\documentclass{aa}  

\usepackage{graphicx}
\usepackage{cancel}
\usepackage{subfig}
\usepackage{txfonts}

\newcommand{\Rsolar}{\mbox{\,$\rm R_{\odot}$}}        
  
\begin{document}

   \title{Asymmetric expansion of coronal mass ejections in the low corona}

   \subtitle{}

   \author{H. Cremades\inst{1,2}
          \and
          F. A. Iglesias\inst{1,2}
          \and
          L.A. Merenda\inst{1}
          }

   \institute{Universidad Tecnol\'ogica Nacional, Facultad Regional Mendoza,
              Centro de Estudios para el Desarrollo Sustentable,
              Rodriguez 243, M5502AJE Mendoza, Argentina\\
              \email{hebe.cremades@frm.utn.edu.ar}
         \and
             Consejo Nacional de Investigaciones Cient\'ificas y T\'ecnicas,
             Godoy Cruz 2290, C1425FQB Ciudad Aut\'onoma de Buenos Aires, Argentina
         }

   \date{Received --; accepted --}

 
  \abstract
   {}
   {Understanding how magnetic fields are structured within coronal mass ejections (CMEs), and how they evolve from the low corona into the heliosphere, is a major challenge for space weather forecasting and for solar physics. The study of CME morphology is a particularly auspicious approach to this problem, given that it holds a close relationship with the CME magnetic field configuration. Although earlier studies have suggested an asymmetry in the width of CMEs in orthogonal directions, this has not been inspected using multi-viewpoint observations.
   }
   {The improved spatial, temporal, and spectral resolution, added to the multiple vantage points offered by missions of the Heliophysics System Observatory, constitute a unique opportunity to gain insight into this regard. We inspect the early evolution (below ten solar radii) of the morphology of a dozen CMEs occurring under specific conditions of observing spacecraft location and CME trajectory, favorable to reduce uncertainties typically involved in the 3D reconstruction used here. These events are carefully reconstructed by means of a forward modeling tool using simultaneous observations of the Solar-Terrestrial Relations Observatory (STEREO) Extreme Ultraviolet Imager (EUVI) and the Solar Dynamics Observatory (SDO) Atmospheric Imaging Assembly (AIA) as input when originating low in the corona, and followed up in the outer fields of view of the STEREO and the Solar and Heliospheric Observatory (SOHO) coronagraphs. We then examine the height evolution of the morphological parameters arising from the reconstructions. 
   }
   {The multi-viewpoint analysis of this set of CMEs revealed that their initial expansion --below three solar radii-- is considerably asymmetric and non-self-similar. Both angular widths, namely along the main axes of CMEs ($AW_L$) and in the orthogonal direction ($AW_D$, representative of the flux rope diameter), exhibit much steeper change rates below this height, with the growth rate of $AW_L$ found to be larger than that of $AW_D$, also below that height. Angular widths along the main axes of CMEs are on average $\approx$\,1.8 times larger than widths in the orthogonal direction $AW_D$. The ratios of the two expansion speeds, namely in the directions of CMEs main axes and in their orthogonal, are nearly constant in time after $\sim$\,4 solar radii, with an average ratio $\approx$\,1.6. Heights at which the width change rate is defined to stabilize are greater for $AW_L$ than for $AW_D$.
   }
   {}

   \maketitle
%

\section{Introduction}

Coronal mass ejections (CMEs) are bright outward-moving structures ejected from the Sun in all directions, detected in the white-light corona by means of coronagraphs. It is well established that magnetic fields associated with interplanetary CMEs, particularly those involving a southward component, can interact with Earth's magnetosphere under certain conditions.  Furthermore, from in situ detections of interplanetary CMEs it is known that there are particular cases, dubbed magnetic clouds (MCs), in which magnetic fields are organized following a helical structure \citep[e.g.,][]{Bothmer-Schwenn1998}. As a consequence, from the practical point of view of space weather, there have been a number of recent efforts devoted to understanding how magnetic fields within CMEs are organized. This necessity has inspired several pieces of work. For instance, \citet{Yurchyshyn-etal2007} looked for a correspondence between the orientation angle of elongated CMEs and of the axis angles of MCs. \citet{Marubashi-etal2015} investigated the connection between the orientation and handedness of interplanetary magnetic flux ropes detected in situ and those of their solar source regions along magnetic polarity inversion lines. Following a similar reasoning,  \citet{Savani-etal2015} proposed a means of predicting the magnetic field structure of interplanetary CMEs arriving at Earth on the basis of their solar sources and the early characteristics of CMEs. 

The study of CME morphology is of particular interest because it holds a close relationship with the magnetic field configuration. Both morphology and its evolution from the earliest stages of CMEs are key factors in order to associate their structure with that of their source region fields. In turn, these two factors  can give hints on the involved initiation processes, and can benefit space weather forecasts on the basis of pre-eruptive field topologies in the low corona or developing CMEs in the corona. 

A fundamental challenge in understanding the morphology of CMEs resides in the limitations imposed by coronagraphic observations. The coronagraph, the prime instrument that enables us to visualize CMEs, is capable of detecting the Thomson scattering brightness produced by coronal electrons, which are shaped by the prevailing magnetic fields. Despite the dramatic improvement in sensitivity and spatial resolution in the past few decades, a fundamental restriction persists: coronagraphs record the integrated brightness along the line of sight, and thus provide a 2D view of a 3D entity, as are CMEs.

\begin{figure}
    \centering
    \includegraphics[width=44mm]{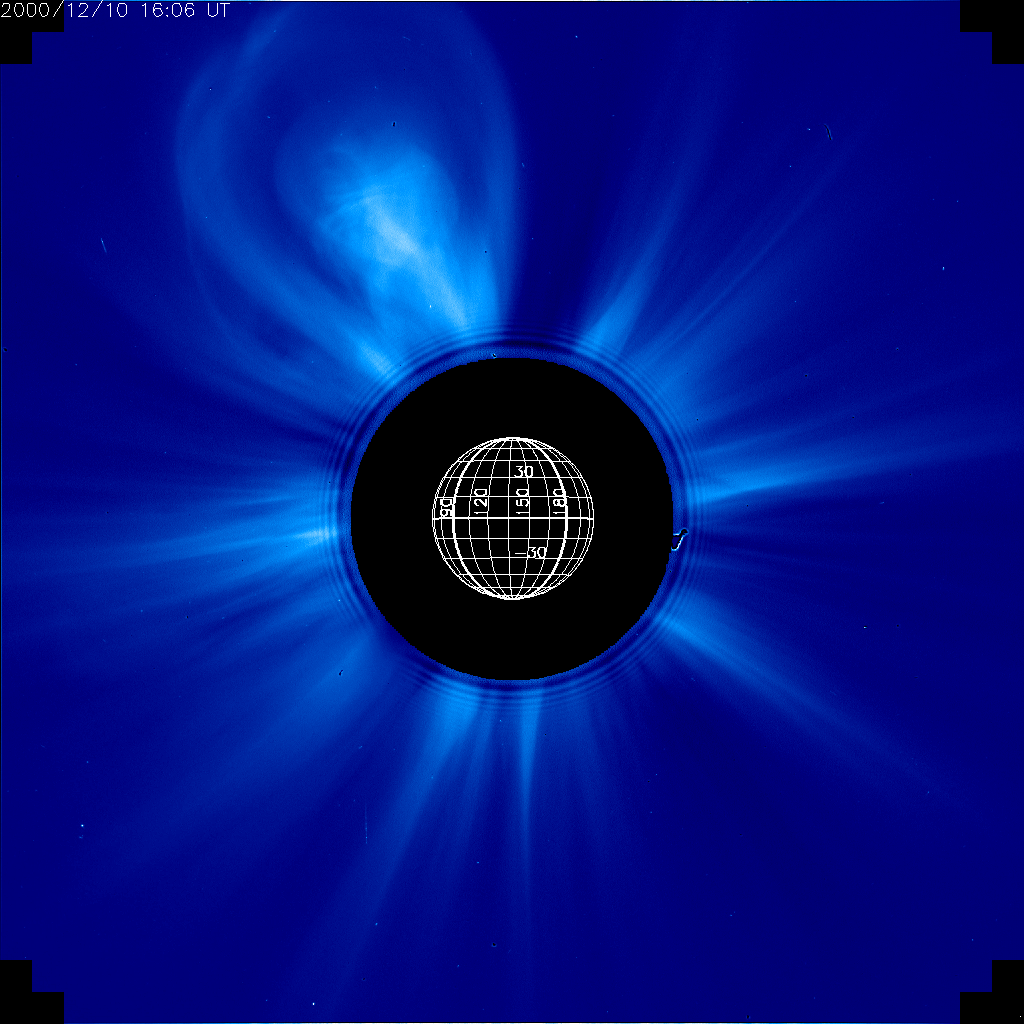}
    \includegraphics[width=44mm]{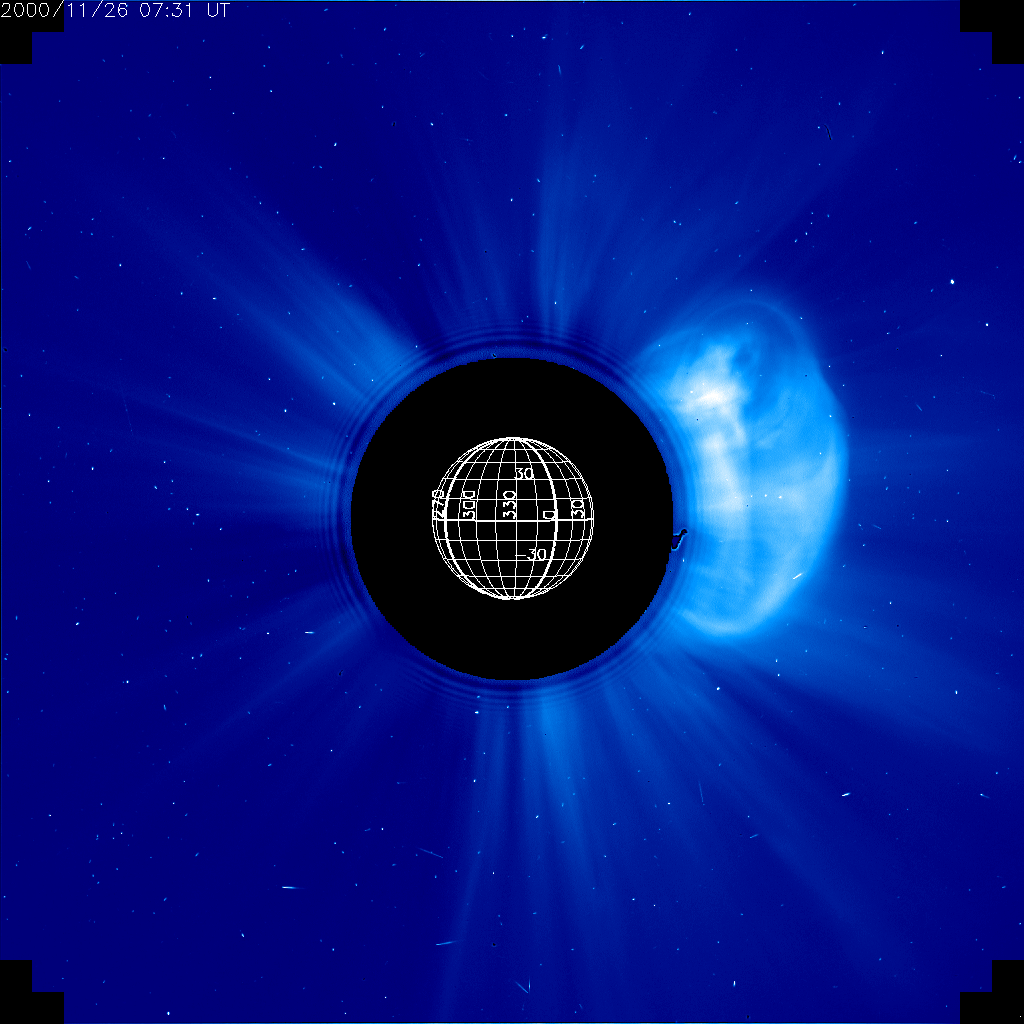}
    \caption{Two CMEs exhibiting extreme cases of projection: axial view, with the LOS aligned with the CME main axis (left); and lateral view, with the LOS nearly perpendicular to the main axis (right).}
    \label{fig:bothtypes}
\end{figure}

Opposite to the early general belief that CMEs were spherical bubbles holding rotational symmetry \citep[e.g.,][]{Crifo-etal1983} supported by observations of Earth-directed circular halo CMEs \citep[e.g.,][]{Howard-etal1982}, a number of studies have indicated that many CMEs \citep[if not all, see ][]{Vourlidas-etal2013}  are organized along a major axis. Moreover, they present a topology consistent with that of a twisted magnetic flux rope in agreement with in situ detections of MCs. Evidence of helical magnetic fields can be found in several CMEs \citep[e.g.,][]{Dere-etal1999, Wood-etal1999, Krall-Chen2001}, while a configuration of arcades with cylindrical symmetry has also been proposed as consistent with CME observations \citep{Moran-Davila2004}. 
\citet{Cremades-Bothmer2004} noted that the projected views of a set of highly structured CMEs depended on the location and inclination of the neutral lines of their associated source regions, also suggestive of a helical flux rope configuration consistent with a cylindrical symmetry. According to this picture, the differing appearances of CMEs may be partly the result of different orientations of their main axis of symmetry with respect to the observer's line of sight (LOS). This configuration leads to two extreme projected views:  {axial}, detected when the LOS is aligned with the main axis of the cylinder, and  {lateral}, seen when the LOS is perpendicular to this axis. These views have been  also referred to respectively as edge-on and face-on views in the literature \citep{Thernisien-etal2006,Thernisien-etal2009,Thernisien-etal2011}, and also axial and broadside \citep[e.g.,][]{Krall-StCyr2006}. The axial (edge-on) view enables detection of the archetypal three-part structure, as in the left panel of Fig. \ref{fig:bothtypes}, with the dark void at times being outlined by circular threads. The lateral (face-on; broadside) view, however, exhibits a more elongated but ragged and diffuse structure (see right panel of Fig. \ref{fig:bothtypes}). Such cases were investigated by \citet{Cremades-Bothmer2005} from the Earth-perspective images provided by the Large Angle Spectroscopic Coronagraph  \citep[LASCO;][]{Brueckner-etal1995} on board SOHO \citep[Solar and Heliospheric Observatory,][]{Domingo-etal1995}. They found that CMEs exhibiting their axial view in coronagraph images were in general much narrower than those exhibiting their lateral perspective. 

Until the mid-2000s, the only feasible view of the solar corona was that provided by instruments on the Sun-Earth line, thus hindering visualization of the same CME from different perspectives. This changed with the STEREO Mission \citep[Solar-Terrestrial Relations Observatory,][]{Kaiser-etal2008}, whose twin spacecraft enabled simultaneous imaging of the solar corona from multiple viewpoints. In fact, a proper and uncommon combination of factors, namely spacecraft separation, CME trajectory, and main axis orientation, enabled the first report of the simultaneous observation of the axial and lateral views of the same CME \citep{Cabello-etal2016}. As pointed out by \citet{Mierla-etal2009}, the three-dimensional (3D) reconstruction of CME morphology from available coronagraphic images is an intrinsically ill-posed problem, given that a proper tomographic reconstruction would require a large number of images of the same CME from numerous different viewpoints. Although this is clearly a limitation to the 3D reconstruction of CMEs, images of the solar corona provided by the twin STEREO spacecraft, in combination with the terrestrial view provided by SOHO, constitute a triplet that better estimates the 3D morphology of CMEs, enabling the study of its evolution from their initiation in the low corona. 

Inspired by the results of \citet{Cremades-Bothmer2004, Cremades-Bothmer2005} and \citet{Cabello-etal2016} concerning the asymmetry in the axial versus lateral perspectives of CMEs, we  examined the early evolution of the morphology of a distinct set of CMEs. The particular conditions of spacecraft location and CME trajectory under which the analyzed events occur reduce uncertainties typically involved in the 3D reconstruction via forward modeling. In Sect. \ref{s:data} we explain why and how these events were analyzed. Section \ref{s:FM} describes the method used to reconstruct the overall 3D morphology of the CMEs under study. Section \ref{s:results} presents the main findings, while Sect. \ref{s:discussion} puts them in the context of previous results and discusses their implications.

\section{Methodology}
\label{s:methods}
\subsection{Data and event selection}
\label{s:data}


White-light coronal images captured from multiple vantage points constitute the main data set used in this analysis. The Earth view is provided by LASCO C2 on board  the legendary SOHO Mission, while the other perspectives are given by the COR1 and COR2 coronagraphs of the STEREO/SECCHI instrument suite \citep[Sun-Earth Connection Coronal and Heliospheric Investigation,][]{Howard-etal2008}. The early evolution of CMEs in the low corona was investigated by means of images in the extreme-ultraviolet (EUV) provided by the Atmospheric
Imaging Assembly  \citep [AIA;][]{Lemen-etal2012} on board the Solar
Dynamics Observatory  \citep[SDO;][]{Pesnell-etal2012} and by the SECCHI Extreme-Ultraviolet Imagers (EUVI) on board STEREO. 

To minimize uncertainties due to projection effects, and with the fundamental goal of characterizing the morphology of a number of CMEs, we  selected events according to the  criteria described below. Otherwise, three views of the same event may not have been sufficient to correctly discern the orientation of its main axis of symmetry or the overall configuration of its magnetic field. First of all, the events under study were identified during times of spacecraft quadrature: November 2010\,--\,July 2011 (STEREO--SOHO quadrature) and December 2012\,--\,June 2013 (quadrature between the two STEREO spacecraft). 

The next and most crucial criterion refers to the direction of propagation of the CMEs, which ideally should be  perpendicular to the plane containing the observing spacecraft (STEREO, SOHO, and SDO) and thus to the ecliptic. Given that most CMEs originate at the two activity belts, those propagating at high latitudes are relatively rare. The reason for this requirement regarding the propagation direction has been addressed in detail by \citet{Cabello-etal2016}. In brief, three distinct views of the same CME are not enough to constrain the morphological parameters if all vantage points and the propagation direction lie on the same plane. This also agrees with the findings of \citet{Wood-etal2009}, who noted that the best results from forward modeling (see Sect. \ref{s:FM}) are obtained when the CME's appearance is very distinct from the different viewpoints, and analogous to the remark from \citet{Thernisien-etal2009}, who noted that imaging instruments placed away from the ecliptic plane would also help resolve ambiguities in the determination of CME parameters.

\begin{figure}
    \sidecaption
    \includegraphics[width=9cm, trim=0cm 18.3cm 0cm 0cm, clip] {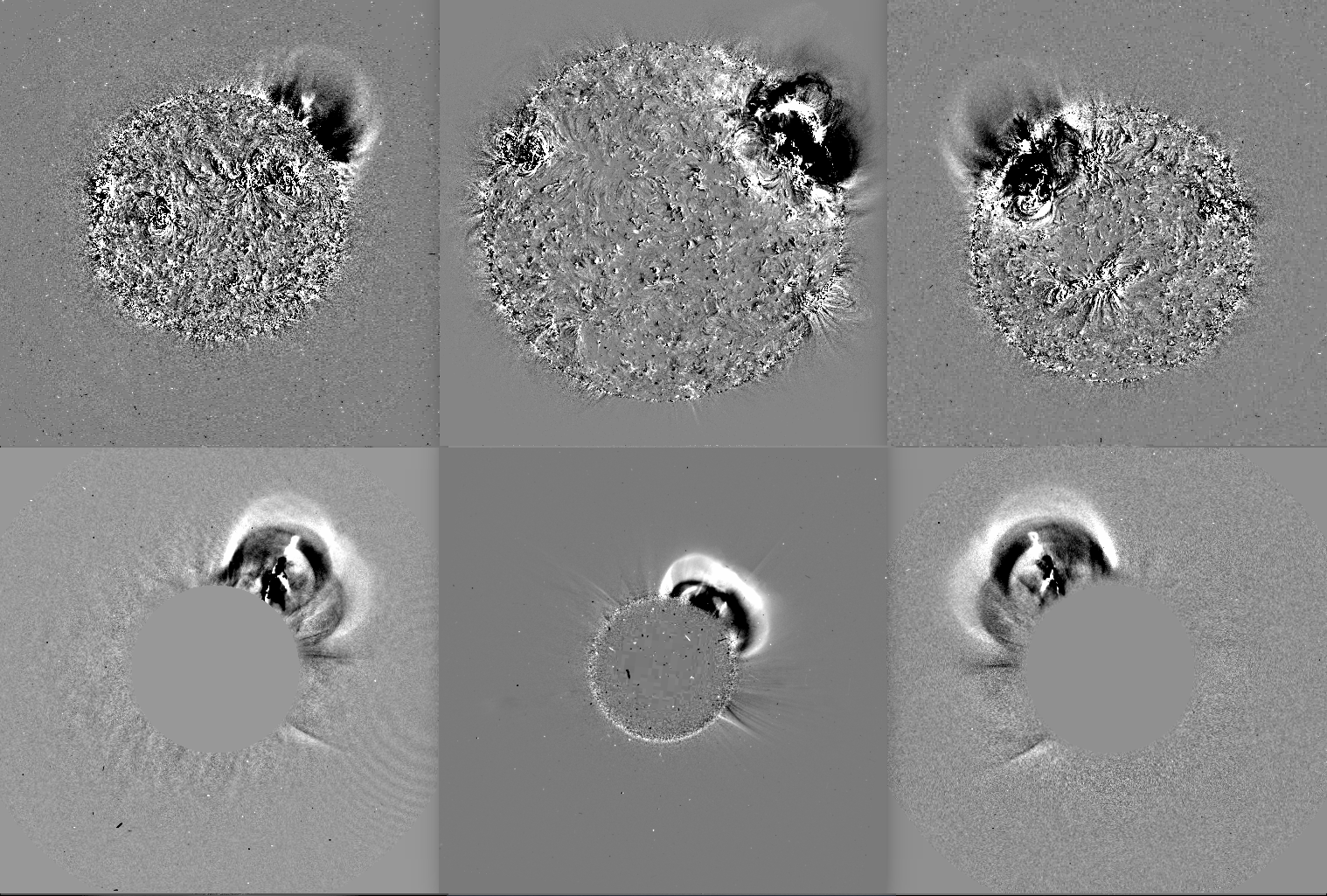}
    \caption{Incipient CME on 14 December 2010 at 15:20 as seen in the low corona from STEREO EUVI-B 195\,\AA~(left), SDO AIA 193\,\AA~(center), and STEREO EUVI-A 195\,\AA~(right). A previous image has been subtracted in all cases to enhance visibility of the event.}
    \label{fig:sampleevent}
\end{figure}

As a first approach to identifying events meeting these criteria, the SOHO LASCO CME Catalog\footnote{\url{http://cdaw.gsfc.nasa.gov/CME_list}} \citep{Yashiro-etal2004} was inspected for CMEs with central position angle (CPA) within $\pm\,25^{\circ}$ with respect to a CPA\,=\,$0^{\circ}$ (north pole) and a CPA\,=\,$180^{\circ}$ (south pole) during the two mentioned quadrature periods. We found a total of 35 events that met the criteria. However, as one of the fundamental objectives of this work is to study the evolution of the 3D morphology of CMEs from their beginnings in the low corona, only those whose eruption could be visualized by the STEREO/SECCHI EUVI and SDO/AIA instruments were considered, in total  12. One of the selected events at an early stage of its evolution is shown in Fig. \ref{fig:sampleevent}.

\subsection{Determination of 3D parameters}
\label{s:FM}

As mentioned above, the general configuration of CMEs has proven to be in agreement with that of a twisted magnetic flux rope having cylindrical symmetry. The Graduated Cylindrical Shell (GCS) model proposed by \citet{Thernisien-etal2009} is a modeling tool capable of adjusting an ad hoc 3D hollow structure, representative of such a flux rope, to two or three nearly simultaneous views of a CME recorded from different vantage points. This forward model is an upgrade to the version developed by \citet{Thernisien-etal2006}, based on the findings of \citet{Cremades-Bothmer2004}. The ad hoc geometrical surface that is fit to the images looks like a hollow croissant, consisting of two conical legs and a front similar to a section of a torus whose cross section  increases with height. Despite its widespread use, the thoughtful application of the model is not trivial. As mentioned, three views of the same CME may not be enough to unambiguously find the true parameters describing the geometrical figure that best fits the CME, since various solutions may seem appropriate, especially if its propagation direction lies close to the plane containing the observing spacecraft. 

To perform this investigation, nearly simultaneous triplets of images at various points in time describing the evolution of the selected events were fit by means of the GCS model. This means that fits were performed for each point in time when the CME was observed, from its beginning in the low corona (see, e.g., \citealt{Patsourakos-etal2010}) and during its propagation through the field of view (FOV) of the coronagraphs STEREO COR1 and COR2, and SOHO/LASCO C2. Initially, proxy values of latitude ($\theta$), longitude ($\phi$), and tilt ($\gamma$) are set to those of the source region and/or associated H$\alpha$ filament. They are then adjusted by visual inspection on a best educated guess basis together with the three remaining parameters of the model, namely height ($H$), half width ($\alpha$; half of the angle between the axes of the CME conical legs), and aspect ratio ($\kappa$; given by $a(r)/r$ with $a(r)$ being the variable radius of each of the conical legs). 
For each event, the triplets of images corresponding to the first and last analyzed times, respectively in EUVI 195\,\AA--AIA 193\,\AA~and in COR2--LASCO C2, are adjusted first. When performing fits for the times comprised in between, the value of each parameter is allowed to vary between those corresponding to the first and last frames. These parameters are iteratively fit until the best visual match of the CME structure in each image and a smooth time evolution of the fit parameters are found. In the first iterations of the fitting process, some events showed a slight jump in  parameter values when going from the COR1 to the COR2 FOV.  To diminish this effect a compromise fit was performed. Naturally, this may result in the masking of slight changes; however, we are interested in the overall temporal evolution of CME widths. Parameters at the last fitted times usually show little to no variation; however, none of them is forced to be constant at a certain height. Figure \ref{fig:GCSfits} shows GCS model fits for all events superimposed to CME snapshots as viewed from the three perspectives. Images of the white-light corona are running-difference to enhance contrast, where special care was taken to avoid misinterpretations between flux rope and shock.

\begin{figure*}[!h] 
    \centering 
    \setlength{\unitlength}{1cm}
    \begin{picture}(11.8,24)
        \put(0,20){\includegraphics[width=11.8cm]{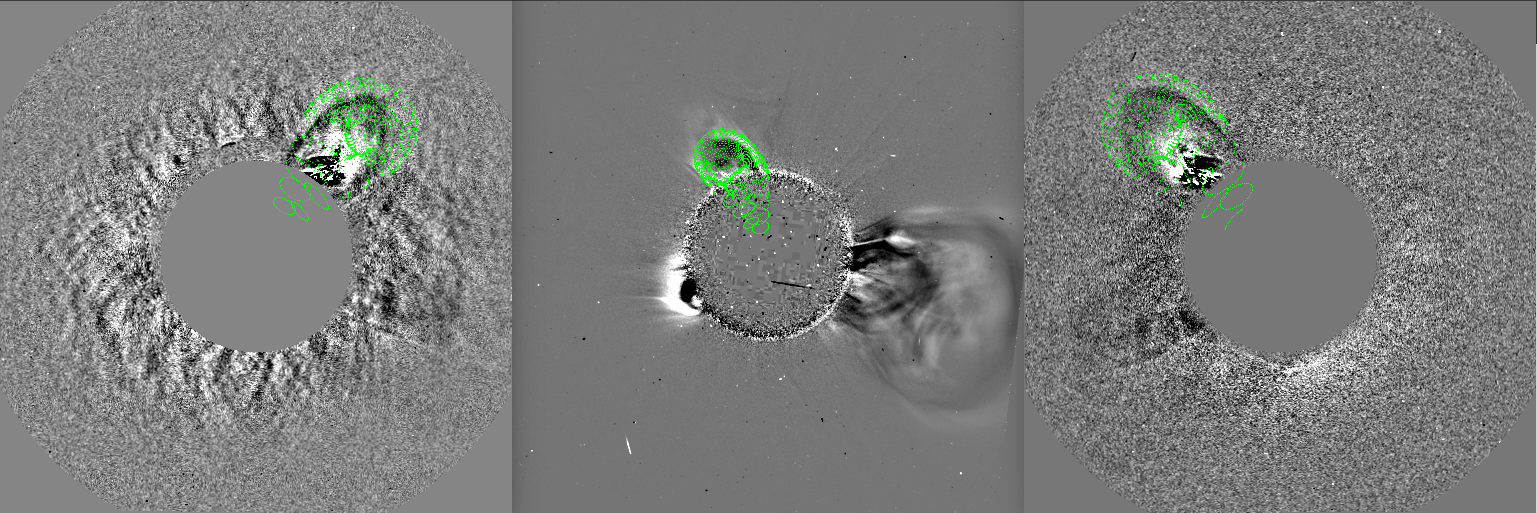}}
        \put(0.1,20.1){\scriptsize{COR2-B 2010/12/12 06:21 UT}}
        \put(4.05,20.1){\scriptsize{LASCO C2 2010/12/12 06:21 UT}}
        \put(8,20.1){\scriptsize{COR2-A 2010/12/12 06:25 UT}}
        
        \put(0,16){\includegraphics[width=11.8cm]{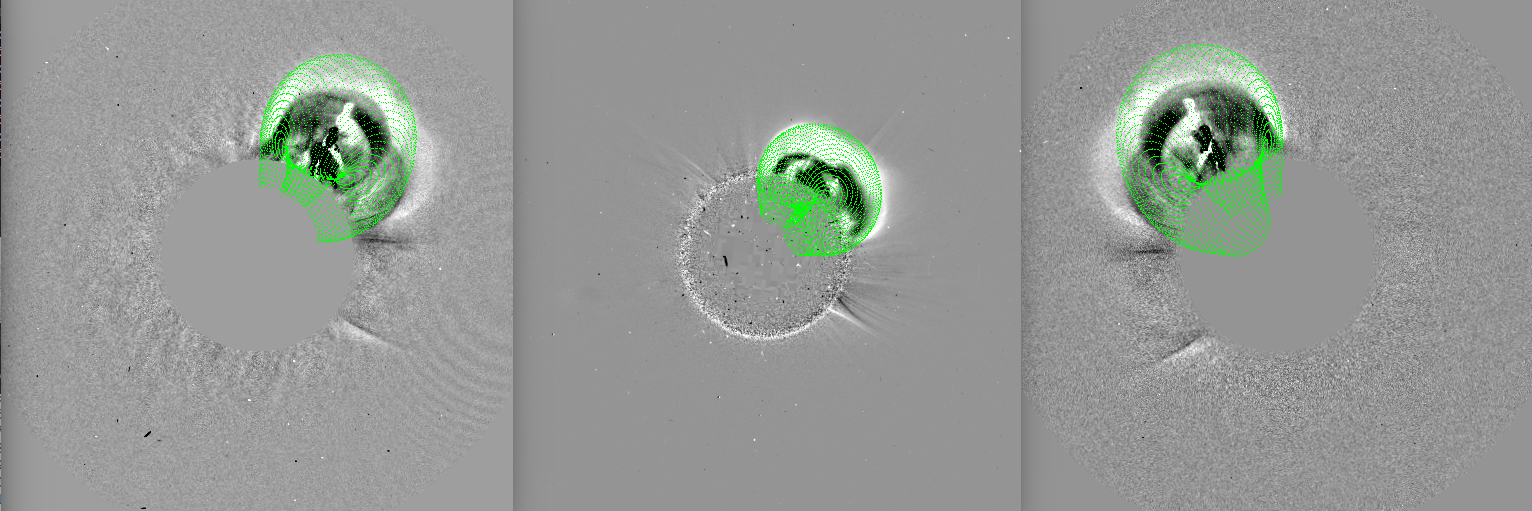}}
        \put(0.1,16.1){\scriptsize{COR2-B 2010/12/14 16:01 UT}}
        \put(4.05,16.1){\scriptsize{LASCO C2 2010/12/14 15:59 UT}}
        \put(8,16.1){\scriptsize{COR2-A 2010/12/14 16:00 UT}}
        
        \put(0,12){\includegraphics[width=11.8cm]{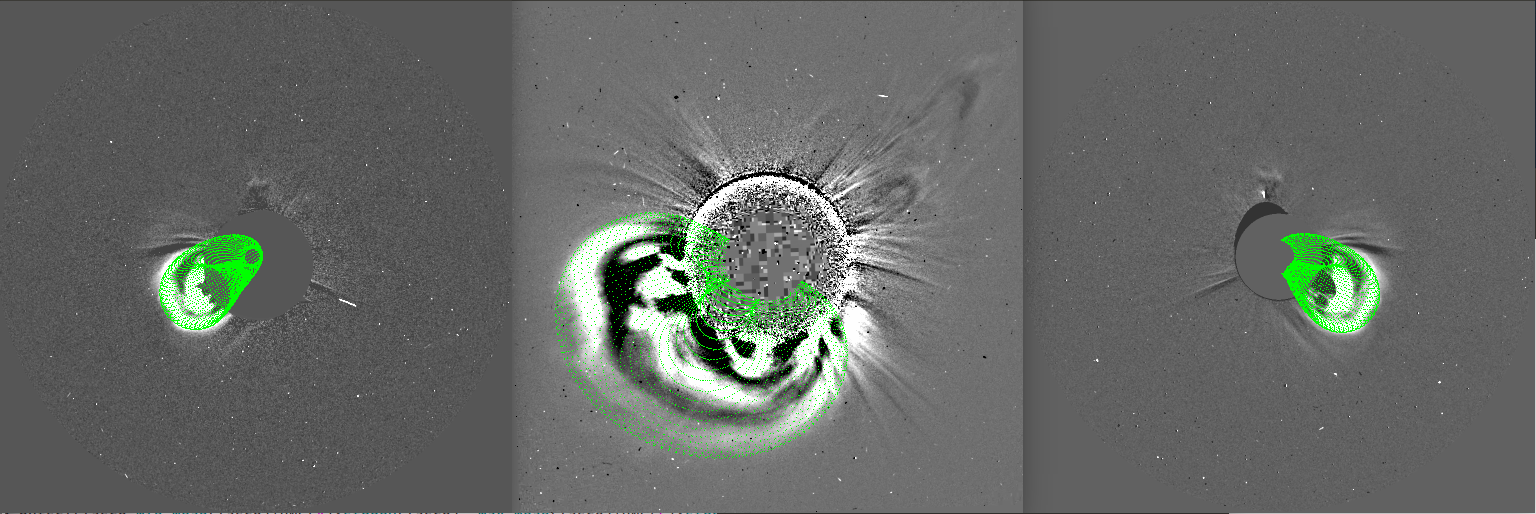}}
        \put(0.1,12.1){\scriptsize{COR2-B 2011/03/17 13:39 UT}}
        \put(4.05,12.1){\scriptsize{LASCO C2 2011/03/17 13:36 UT}}
        \put(8,12.1){\scriptsize{COR2-A 2011/03/17 13:39 UT}}
        
        \put(0,8){\includegraphics[width=11.8cm]{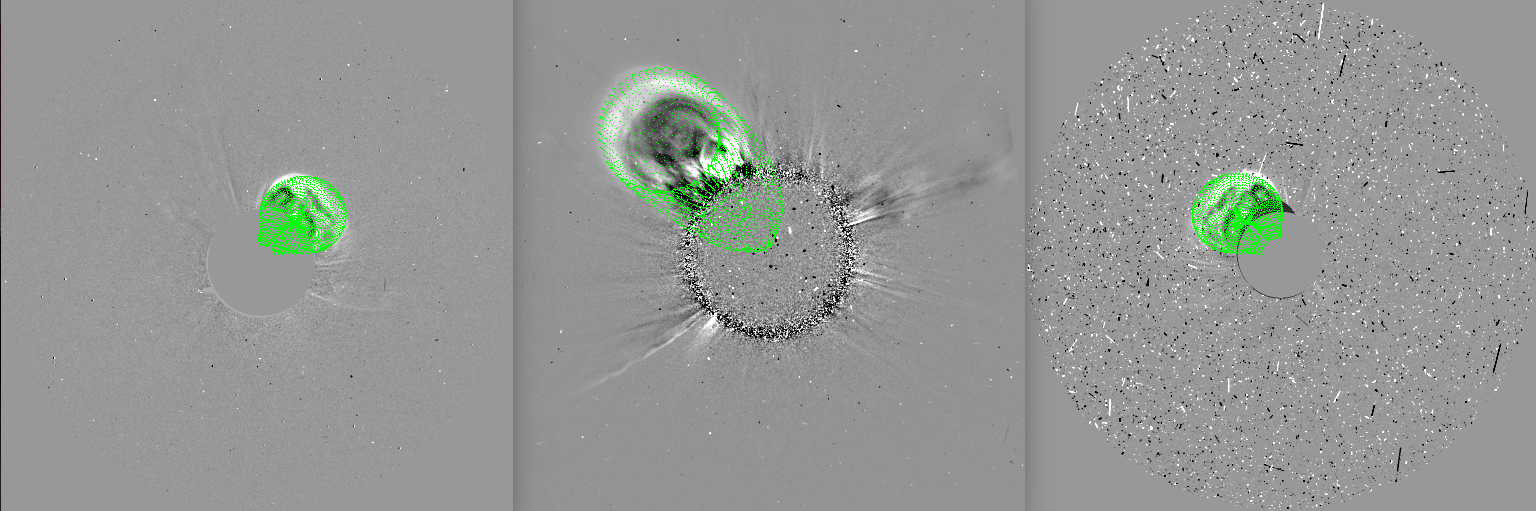}}
        \put(0.1,8.1){\scriptsize{COR2-B 2011/06/05 07:54 UT}}
        \put(4.05,8.1){\scriptsize{LASCO C2 2011/06/05 07:53 UT}}
        \put(8,8.1){\scriptsize{COR2-A 2011/06/05 07:54 UT}}
        
        \put(0,4){\includegraphics[width=11.8cm]{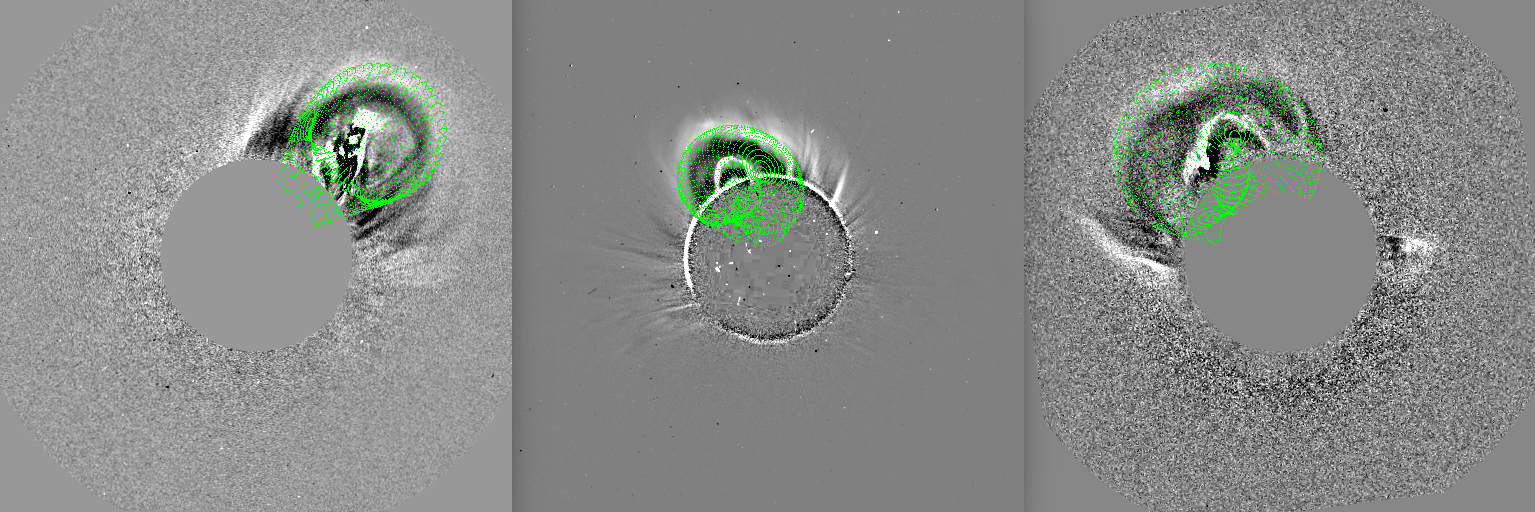}}
        \put(0.1,4.1){\scriptsize{COR2-B 2013/01/23 14:46 UT}}
        \put(4.05,4.1){\scriptsize{LASCO C2 2013/01/23 14:47 UT}}
        \put(8,4.1){\scriptsize{COR2-A 2013/01/23 14:45 UT}}
        
        \put(0,0){\includegraphics[width=11.8cm]{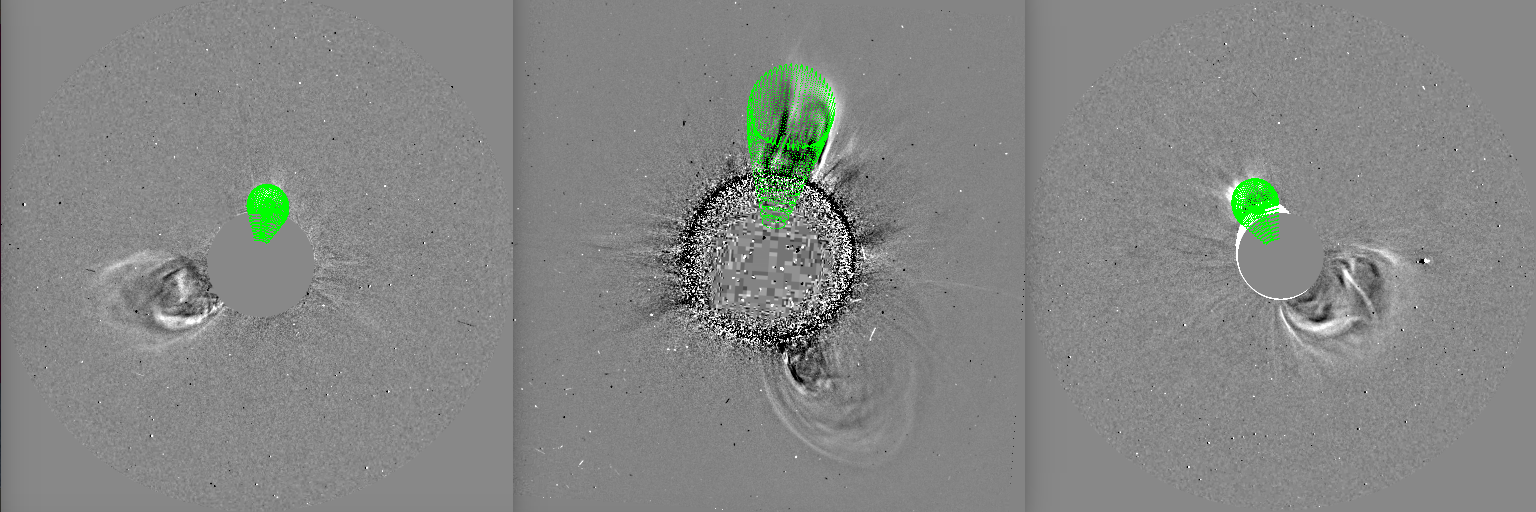}}
        \put(0.1,0.1){\scriptsize{COR2-B 2013/01/29 04:25 UT}} \put(4.05,0.1){\scriptsize{LASCO C2 2013/01/29 04:24 UT}}
        \put(8,0.1){\scriptsize{COR2-A 2013/01/29 04:24 UT}}
    \end{picture}
    \caption{GCS model fits (green meshes) superimposed on snapshots of the 12 analyzed high-latitude CMEs. Each row corresponds to the three views of a particular event, while the left, middle, and right columns are the views from coronagraphs on board STEREO-B, SOHO, and STEREO-A, respectively.}
    \label{fig:GCSfits}
\end{figure*}

\begin{figure*}[!h]
    \ContinuedFloat
    \centering
        \setlength{\unitlength}{1cm}
    \begin{picture}(11.8,24)
        \put(0,20){\includegraphics[width=11.8cm]{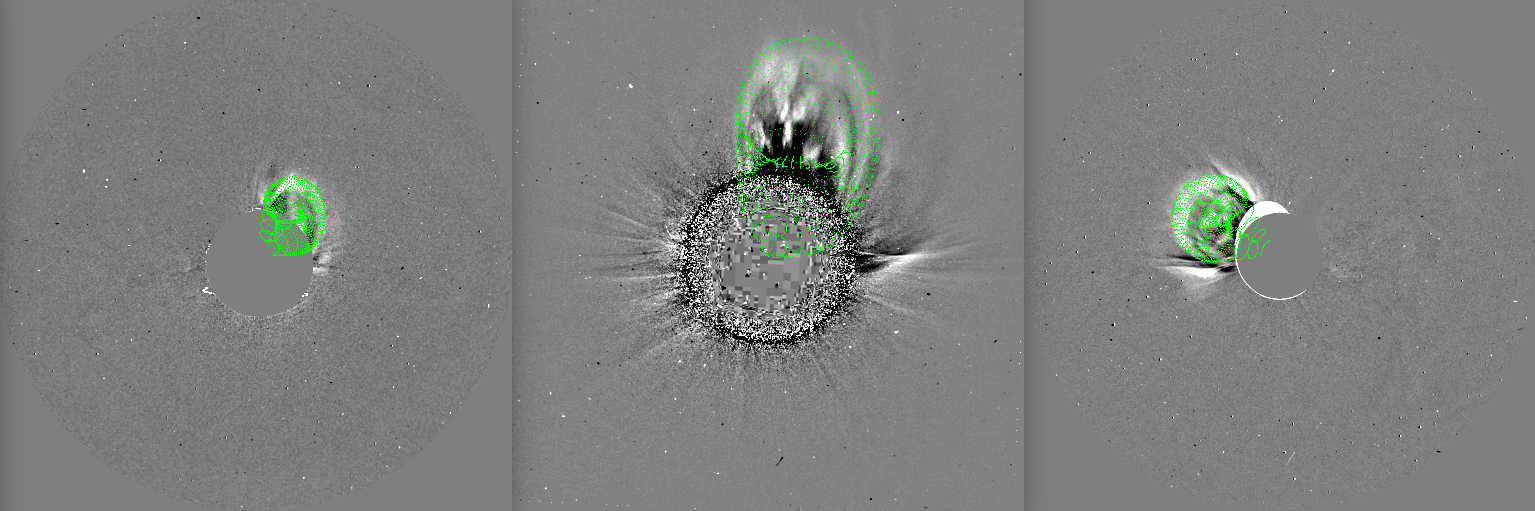}}
        \put(0.1,20.1){\scriptsize{COR2-B 2013/02/09 07:40 UT}}
        \put(4.05,20.1){\scriptsize{LASCO C2 2013/02/09 07:35 UT}}
        \put(8,20.1){\scriptsize{COR2-A 2013/02/09 07:39 UT}}
        
        \put(0,16){\includegraphics[width=11.8cm]{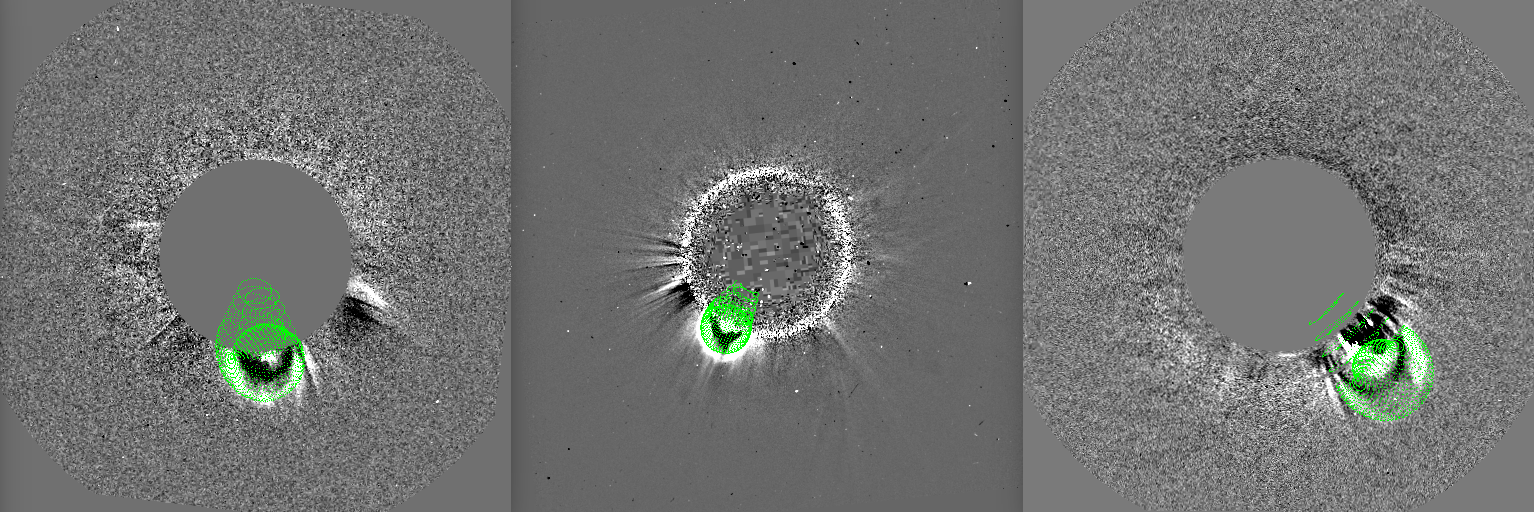}}
        \put(0.1,16.1){\scriptsize{COR2-B 2013/04/24 06:20 UT}}
        \put(4.05,16.1){\scriptsize{LASCO C2 2013/04/24 06:23 UT}}
        \put(8,16.1){\scriptsize{COR2-A 2013/04/24 06:20 UT}}
        
        \put(0,12){\includegraphics[width=11.8cm]{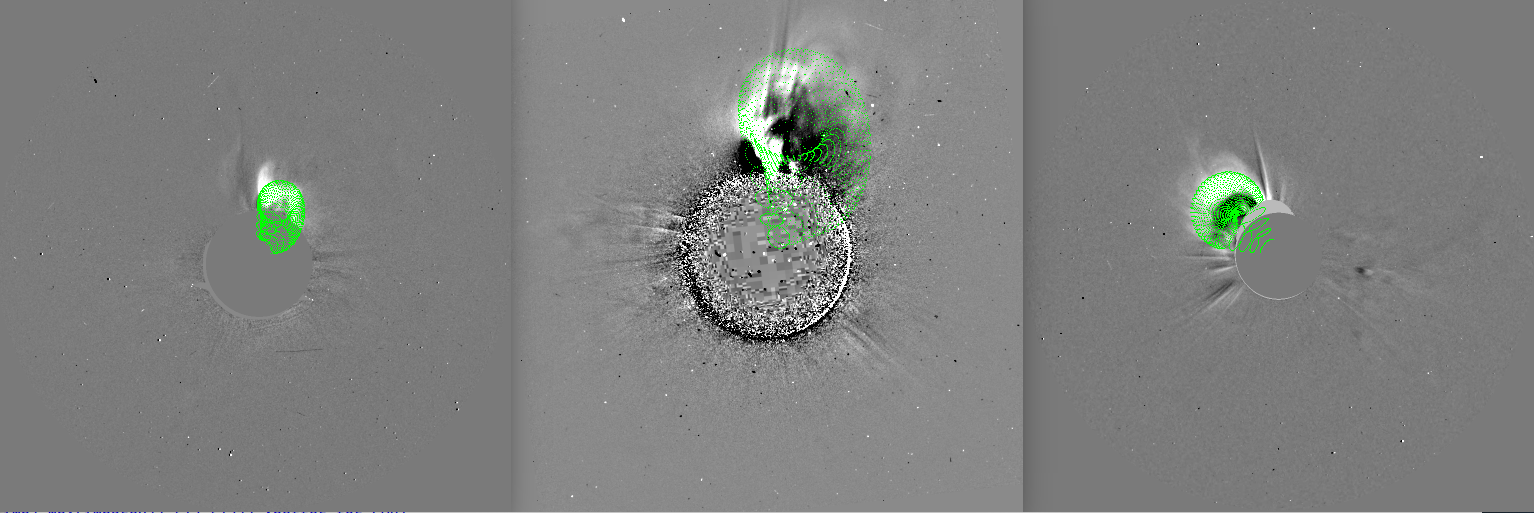}}
        \put(0.1,12.1){\scriptsize{COR2-B 2013/05/02 06:24 UT}}
        \put(4.05,12.1){\scriptsize{LASCO C2 2013/05/02 06:22 UT}}
        \put(8,12.1){\scriptsize{COR2-A 2013/05/02 06:24 UT}}
        
        \put(0,8){\includegraphics[width=11.8cm]{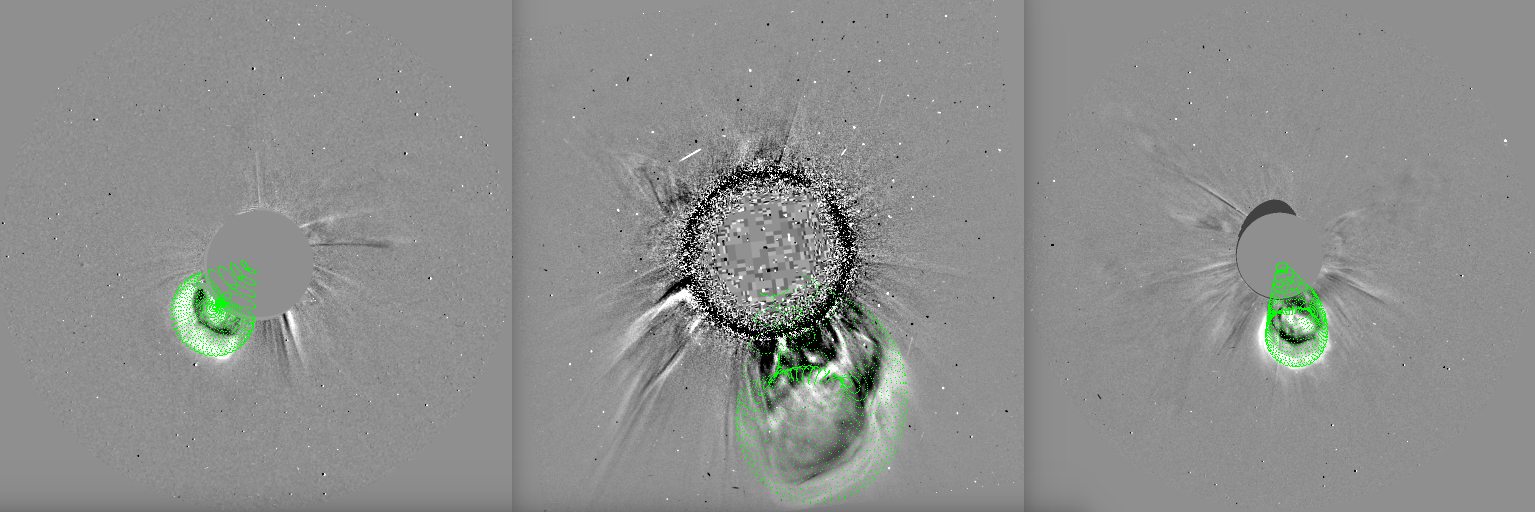}}
        \put(0.1,8.1){\scriptsize{COR2-B 2013/05/17 22:24 UT}}
        \put(4.05,8.1){\scriptsize{LASCO C2 2013/05/17 22:23 UT}}
        \put(8,8.1){\scriptsize{COR2-A 2013/05/17 22:24 UT}}
        
        \put(0,4){\includegraphics[width=11.8cm]{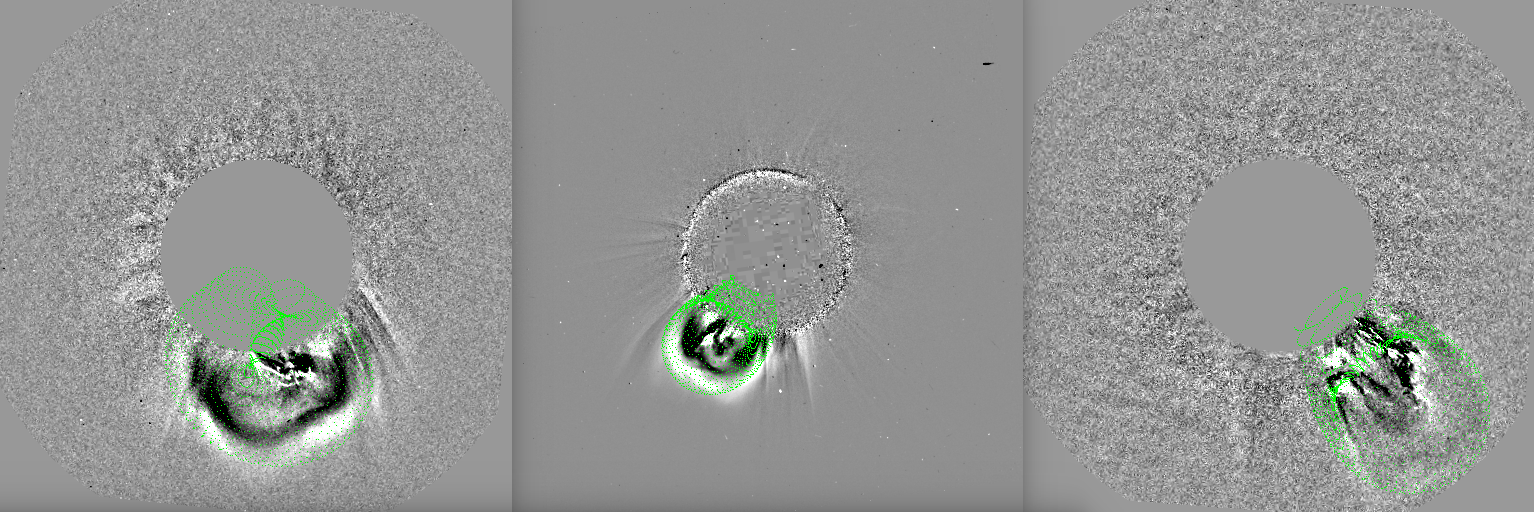}}
        \put(0.1,4.1){\scriptsize{COR2-B 2013/05/27 20:10 UT}}
        \put(4.05,4.1){\scriptsize{LASCO C2 2013/05/27 20:11 UT}}
        \put(8,4.1){\scriptsize{COR2-A 2013/05/27 20:10 UT}}
        
        \put(0,0){\includegraphics[width=11.8cm]{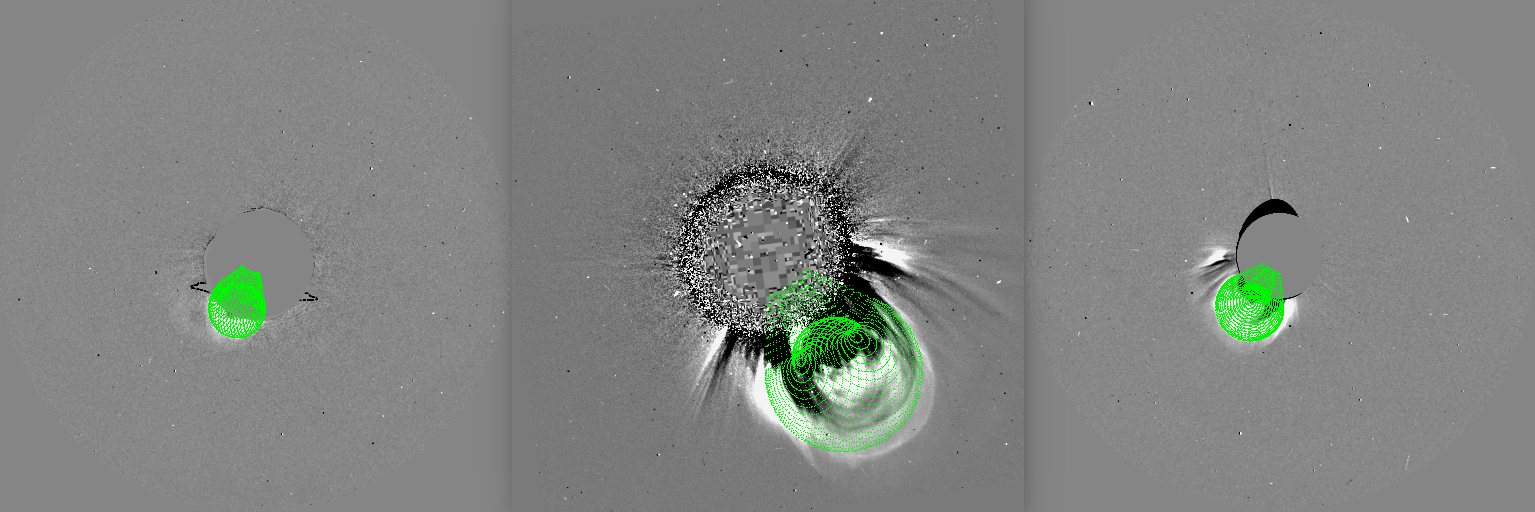}}
        \put(0.1,0.1){\scriptsize{COR2-B 2013/06/07 23:54 UT}} \put(4.05,0.1){\scriptsize{LASCO C2 2013/06/08 00:00 UT}}
        \put(8,0.1){\scriptsize{COR2-A 2013/06/07 23:54 UT}}
    \end{picture}
    \caption{(cont.) GCS model fits (green meshes) superimposed on snapshots of the 12 analyzed high-latitude CMEs.}
\end{figure*}

\begin{figure*}
    \sidecaption
    \includegraphics[width=80mm, angle=90, clip=true]{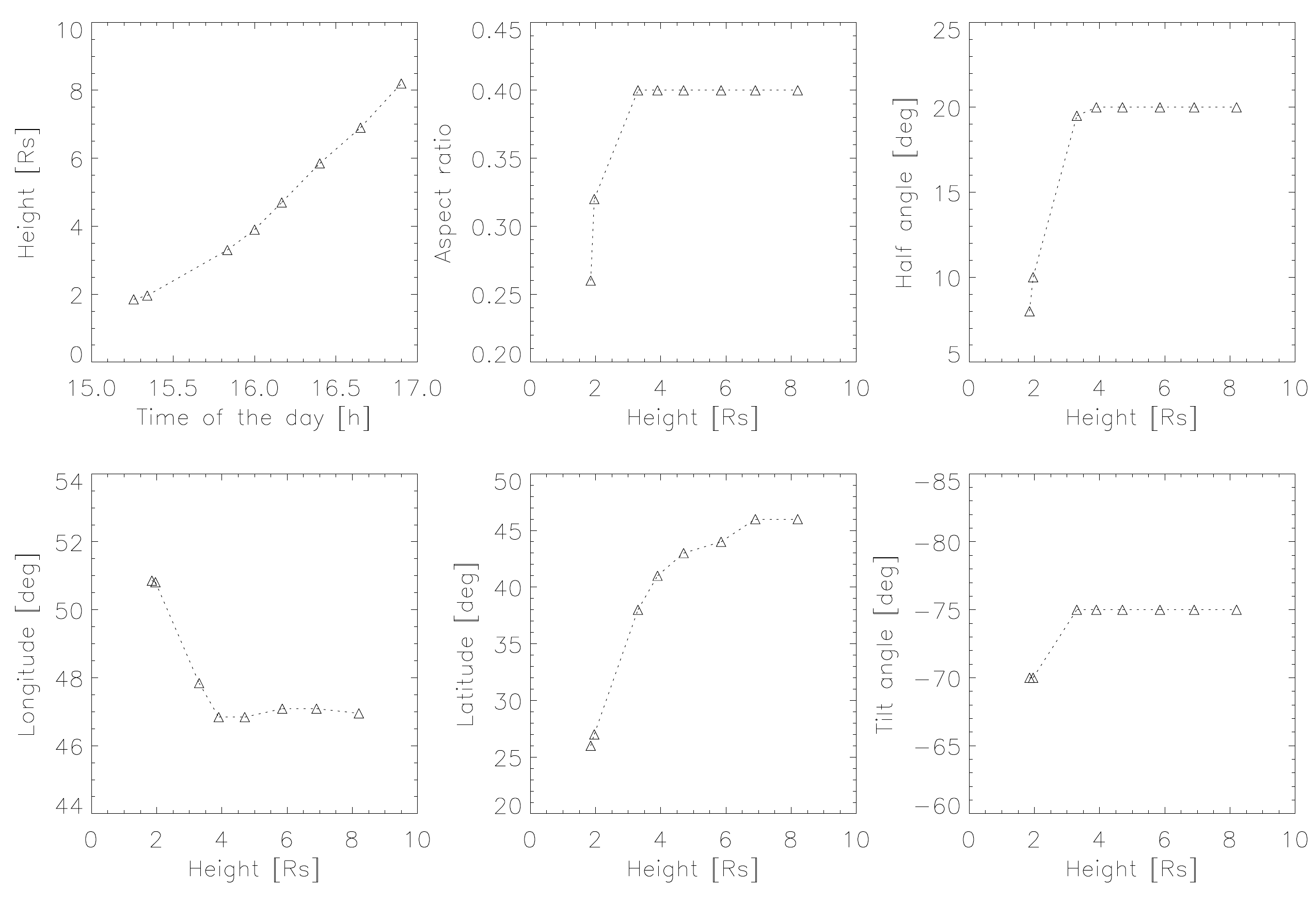}
    \caption{Evolution of the GCS parameters deduced for the analyzed event on 14 December 2010. From left to right and top to bottom: CME height $H$, aspect ratio $\kappa$, half angle $\alpha$, longitude $\phi$, latitude $\theta$, and tilt $\gamma$. The evolution is shown as a function of height in all panels except for the height, which is shown against time.}
    \label{fig:evolution}
\end{figure*}

For each event, plots like those  in Fig. \ref{fig:evolution} were produced to show the evolution of the GCS parameters. CME height $H$ is displayed as a function of time, while all other parameters ($\alpha, \phi, \theta, \kappa, \gamma$) are shown as a function of height. Given that there is no merit function that quantitatively estimates the goodness of the fit, and thus the GCS parameters uncertainties, and that the interpretation of the geometry may differ from event to event and from observer to observer, precise quantitative errors for the fitted GCS parameters cannot be determined \citep[see, e.g.,][]{Mierla-etal2009}, only roughly estimated (see Sect. \ref{s:results}). Moreover, we would like to emphasize that we are more interested in the temporal evolution of CME parameters than in their absolute values.

The main 3D attributes found for the investigated CMEs are summarized in Table \ref{table}. The first column displays the date and time of the first set of reconstructed parameters for each event. The second and third columns refer to the final central latitude and longitude arising from the GCS reconstructions. The fourth and fifth columns show the radial propagation speed at 6\,\Rsolar~and the mean acceleration within the analyzed time interval. The remaining columns will be addressed in the following section.

\begin{figure*}[!h]
    \sidecaption
    \includegraphics[width=9cm, angle=90, trim=0.7cm 2cm 2.0cm 1.7cm, clip=true]{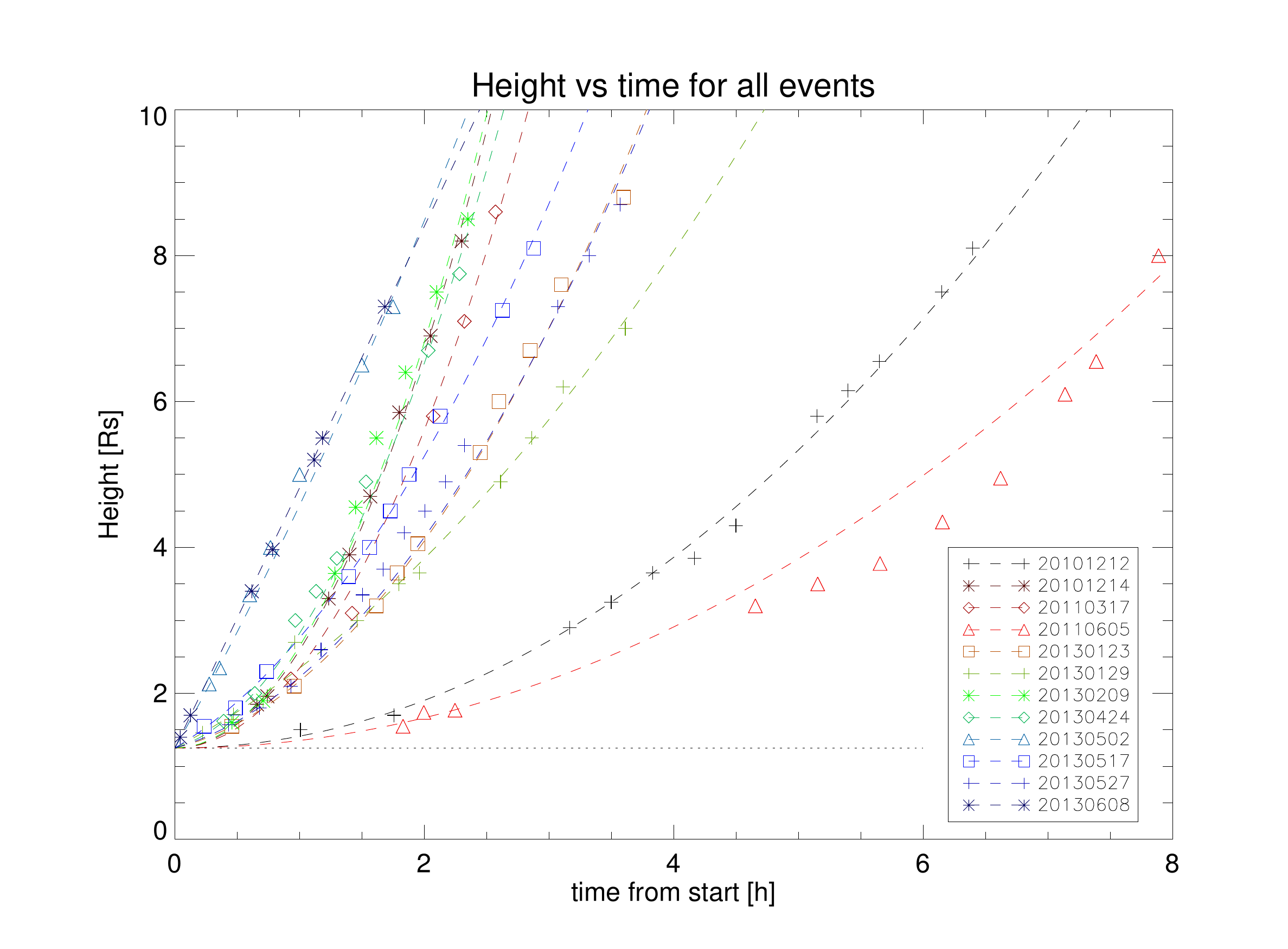}
    \caption{Height-time plots arising from the GCS fits to the 12 analyzed events. The various symbol-color combinations correspond to the different events, and the dashed lines are constrained second-order fits.}
    \label{fig:h-t}
\end{figure*}

\begin{table*}
\centering
\small
\begin{tabular}{ccccccccc}
\hline\hline
Date \& Time & Latitude & Longitude  & Radial Speed & Acceleration & Final AW$_D$ & Final AW$_L$ & Axial Speed & Lateral Speed \\
{[}UT{]} & {[}deg{]} & {[}deg{]} & {[}km s$^{-1}${]} & {[}m s$^{-2}${]} & [deg] & [deg] & {[}km s$^{-1}${]} & {[}km s$^{-1}${]}\\
(1)&(2)&(3)&(4)&(5)&(6)&(7)&(8)&(9)\\
\hline
2010/12/12 03:15:30 &       46 &      -26 &      340 &       17 &       27$\pm$       7 &       51$\pm$      13 & 111 & 190\\
2010/12/14 15:15:30 &       45 &       46 &      984 &      146 &       47$\pm$      13 &       87$\pm$      24 & 670 & 1039\\
2011/03/17 12:15:30 &      -24 &     -149 &      877 &      116 &       53$\pm$      15 &      143$\pm$      38 & 1319 & 2489\\
2011/06/05 02:20:30 &       33 &      -38 &      271 &       11 &       38$\pm$      10 &       94$\pm$      22 & 172 & 321\\
2013/01/23 13:15:30 &       42 &      -12 &      601 &       53 &       47$\pm$      13 &       77$\pm$      21 & 363 & 512\\
2013/01/29 02:00:30 &       63 &       19 &      416 &       21 &       31$\pm$       9 &       54$\pm$      16 & 130 & 188\\
2013/02/09 06:00:30 &       26 &       10 &      991 &      148 &       34$\pm$       9 &       64$\pm$      16 & 442 & 694\\
2013/04/24 05:45:30 &      -49 &     -148 &      874 &      114 &       33$\pm$       9 &       46$\pm$      13 & 303 & 379\\
2013/05/02 05:10:30 &       36 &       13 &      740 &       30 &       31$\pm$       9 &       65$\pm$      18 & 304 & 539\\
2013/05/17 20:00:30 &      -49 &      154 &      620 &       52 &       38$\pm$      10 &       68$\pm$      17 & 330 & 513\\
2013/05/27 18:45:30 &      -49 &     -134 &      587 &       50 &       47$\pm$      12 &       81$\pm$      22 & 358 & 511\\
2013/06/07 22:45:30 &      -61 &       77 &      692 &        1 &       53$\pm$      14 &       70$\pm$      18 & 506 & 619 \\
\hline
\end{tabular}
\caption{Analyzed events and main attributes. Column 1: date and time of the first GCS fit point. Columns 2 and 3: final central latitude and longitude from GCS reconstructions. Columns 4 and 5: radial propagation speed at 6\,\Rsolar~and mean acceleration within the analyzed time interval. Columns 6 and 7: final values of $AW_D$ and $AW_L$ (see next section). Columns 8 and 9: expansion speeds in the axial and lateral directions at 6\,\Rsolar.}
\label{table}
\end{table*}

\section{Results}
\label{s:results}

Since we would like to understand how CMEs expand as they propagate outward in the corona, first we need to examine their height-time behavior. Figure \ref{fig:h-t} displays the height-time plots of the 12 analyzed events, where the symbols represent the data points arising from the GCS fitting. The dashed lines are second-order functions that best fit the time evolution of each event. To normalize time and avoid negative velocities, each fit curve starts at a height of 1.25\,\Rsolar, which is a typical height for quiescent coronal cavities before being triggered to eruption \citep{Gibson-etal2006}.  In this sample, all events exhibit either an accelerated or a linear profile. To follow a smooth general trend, the quantities analyzed below rely on the height values that arise from the second-order fits rather than on the measurements themselves.

\begin{figure}[!h]
    \centering
    \includegraphics[width=6.7cm, angle=90, trim=0.7cm 2cm 2.0cm 1.6cm, clip=true]{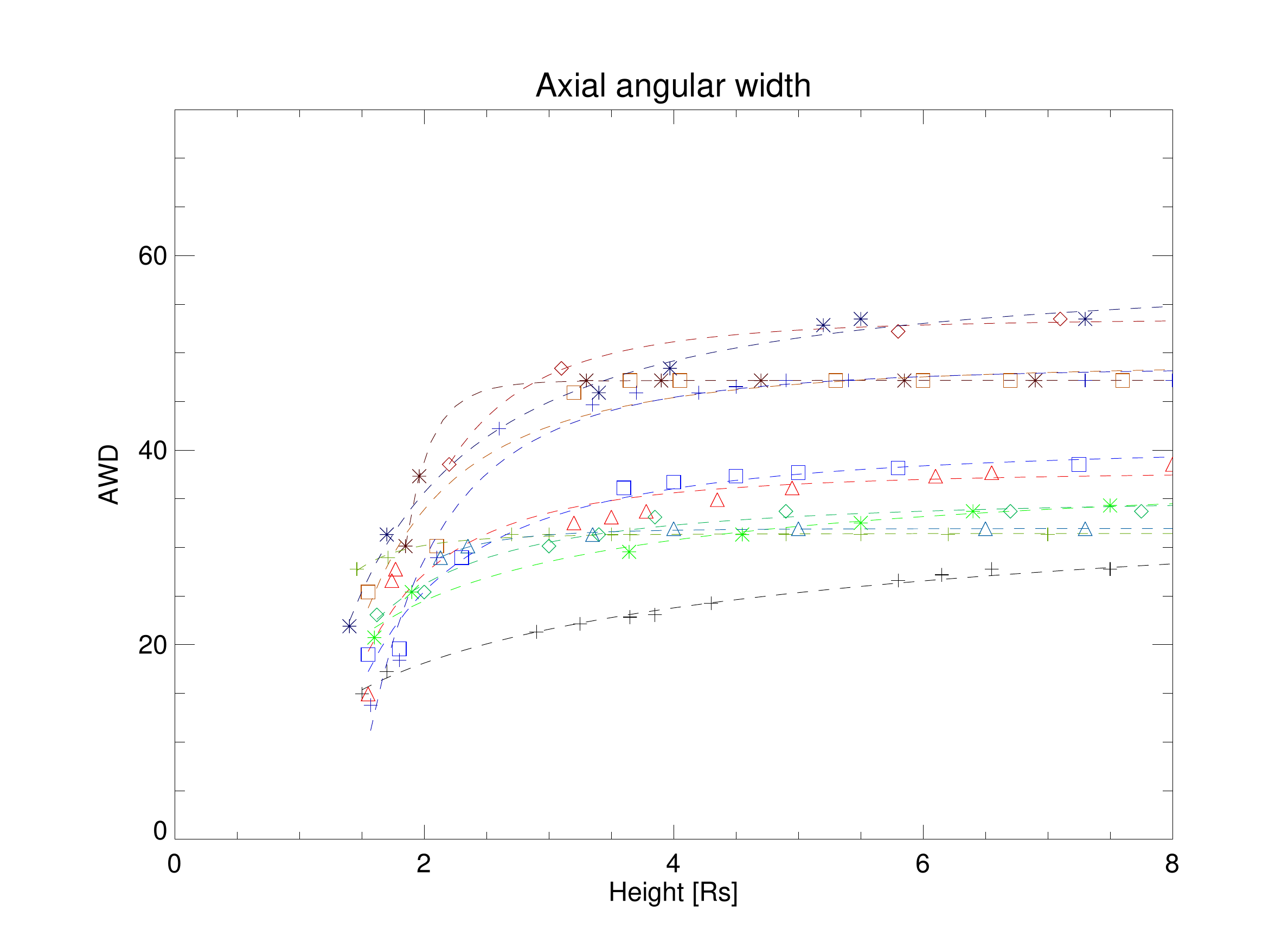}\\
    \vspace{1mm}
    \includegraphics[width=6.7cm, angle=90, trim=0.7cm 2cm 2.0cm 1.6cm, clip=true]{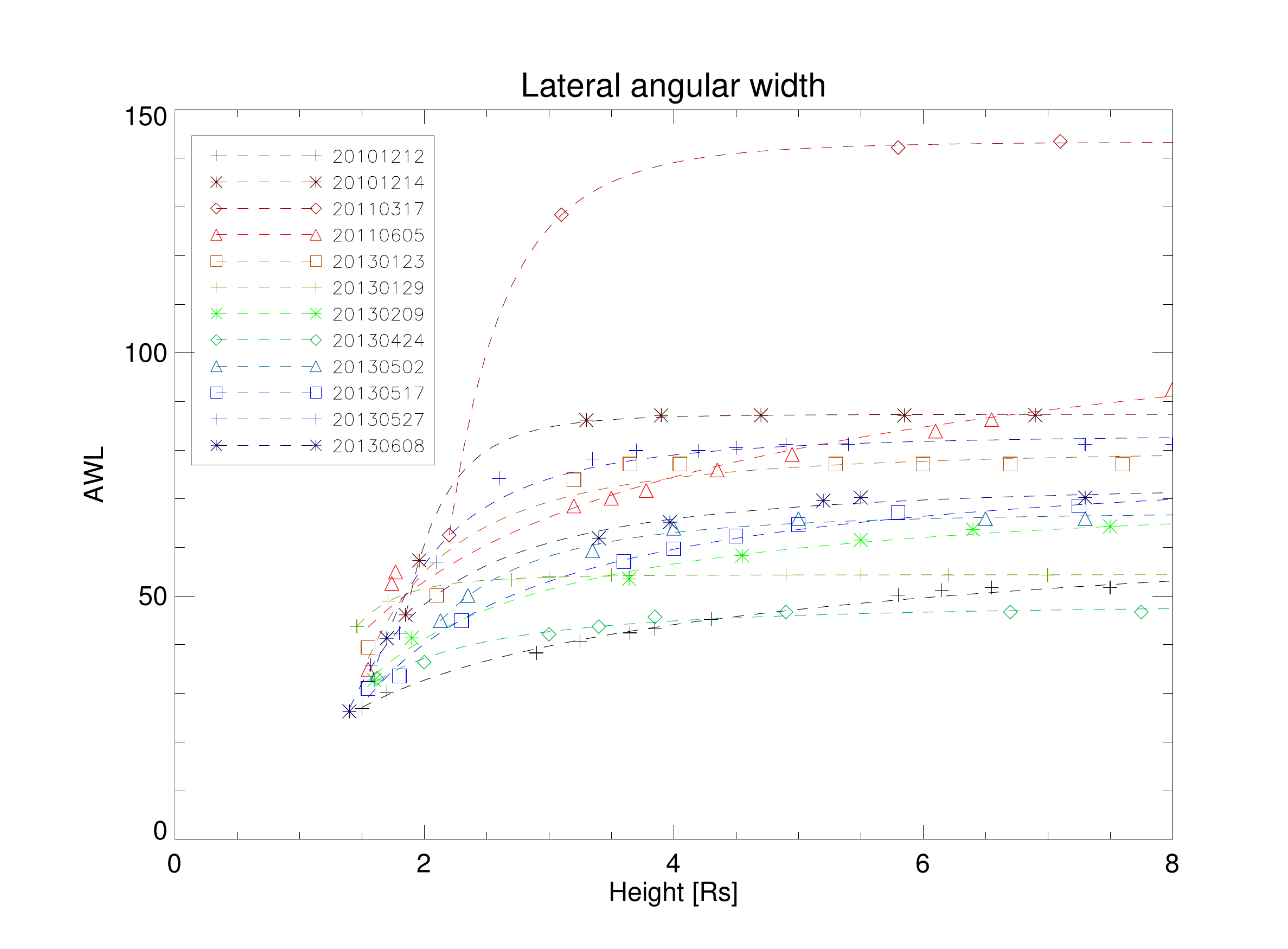}
    \caption{Height evolution of the CME AW in the axial ($AW_D$, top panel) and lateral ($AW_L$, bottom panel) directions. The symbols represent the data points, while the dashed lines are fit power laws extending back to the first data point of each event (see legend in the bottom panel, which also applies to the top plot).}
    \label{fig:bothAWs}
\end{figure}

To understand how CMEs expand in the direction of their main axis and  orthogonal to it, the angular widths in the axial ($AW_D$) and lateral ($AW_L$) directions are of particular interest. They correspond to the views shown in Fig. 1, and are determined as $AW_D = 2\delta$ and $AW_L = 2\alpha + \delta$; with $\delta$ being the $arcsin(\kappa)$ \citep{Thernisien-etal2011}. Both AWs are plotted as a function of height in Fig. \ref{fig:bothAWs}, along with fit power laws extending back to the earliest data point of each event. A rapid increase in the AW in both axial and lateral directions is evident in the first 3\,\Rsolar, followed by a stabilization phase at larger heights. The final values of $AW_D$ and $AW_L$ are listed in  Cols. 6 and 7 of Table \ref{table}, followed by a measure of the corresponding uncertainties arising from the half angle $\alpha$ and the aspect ratio $\kappa$.  These values have been estimated,  as was done by \citet{Patsourakos-etal2010}, by verifying that the fits become unacceptable when varying $\alpha$ and $\kappa$ beyond 20\,\% of the chosen values for the frame at the lowest height. There is no correlation between the final values of $AW_L$ and $AW_D$: events with large final $AW_L$ do not necessarily imply a large final $AW_D$. It is thus worth noting that $AW_L$ and $AW_D$ depend on $\delta$, but only $AW_L$ also depends  on $\alpha$. Since $\alpha$ and $\delta$ are free parameters of the model, $AW_L$ and $AW_D$ can vary independently. The initial change rate (deg\,h$^{-1}$) reaches a few hundred for some events, and drops off  to a few tens of degrees per hour by 3\,\Rsolar. 

\begin{figure*}
    \sidecaption
    \centering
    \includegraphics[width=9cm, angle=90, trim=0.7cm 2cm 1.3cm 1.7cm, clip=true]{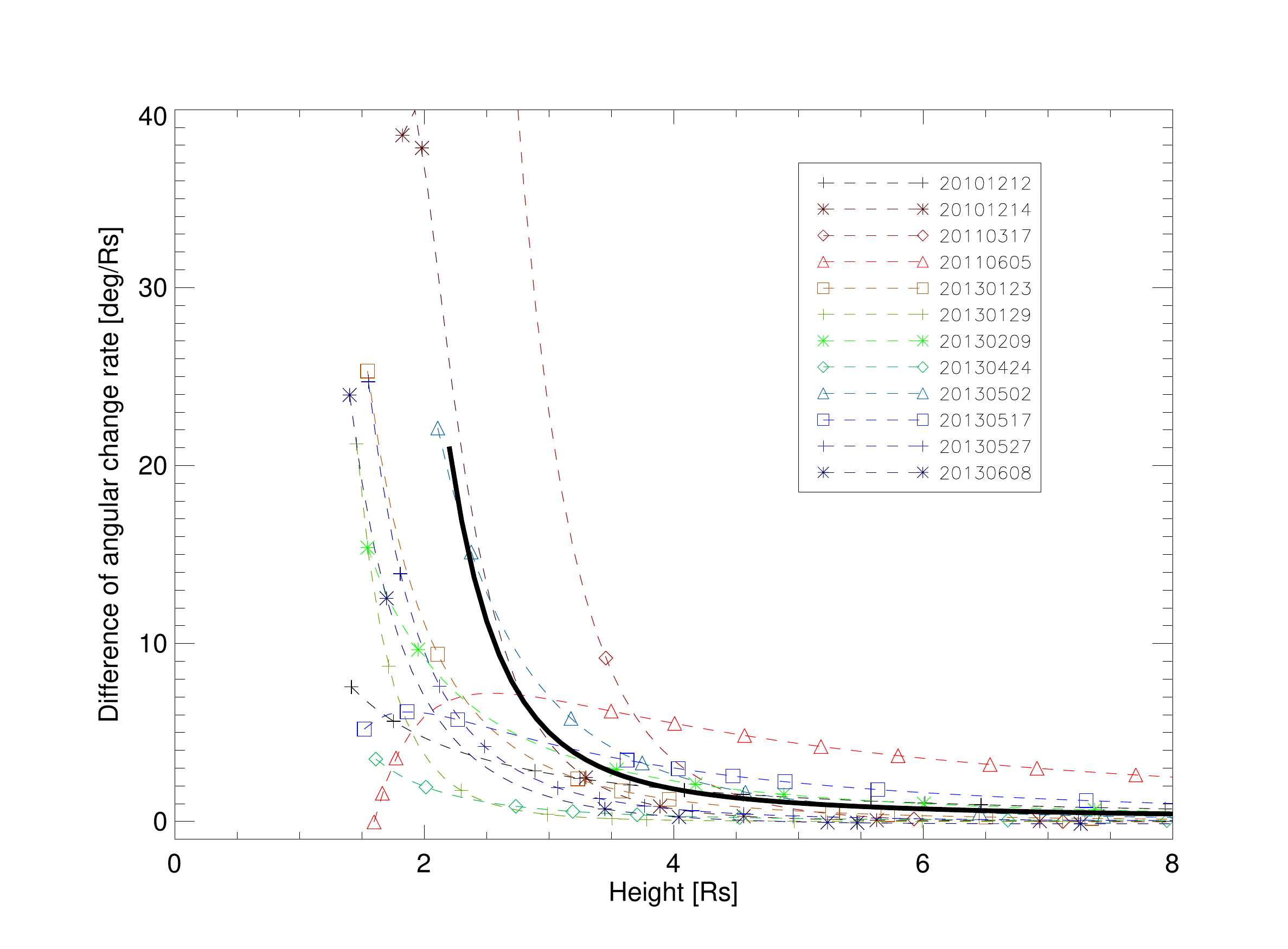}
    \caption{Difference between the fit height change rate of the angular widths in the lateral direction and in the axial direction for each event (dashed lines, see the legend). The black solid line is the average curve, starting at the first height where all events could be measured. The symbols here only indicate the location over the curves where the event was measured. The curve corresponding to 20110317 exceeds the scale and has been intentionally cut off.}
    \label{fig:difference}
\end{figure*}

Figure \ref{fig:difference} shows the difference between the derivatives with respect to height of the fit $AW_L$ and $AW_D$ curves in Fig. \ref{fig:bothAWs}. This difference is always positive, denoting that the AW grows faster in the direction of the main axis of CMEs (lateral direction), than perpendicular to it (axial direction). Moreover, the difference is much larger in the first $\sim$\,3\,\Rsolar, surpassing the 10 degrees per solar radius. After that height both expansion rates and their differences drop to become negligible. 

\begin{figure}
    \centering
    \includegraphics[width=6.7cm, angle=90, trim=0.7cm 2cm 1.3cm 1.7cm, clip=true]{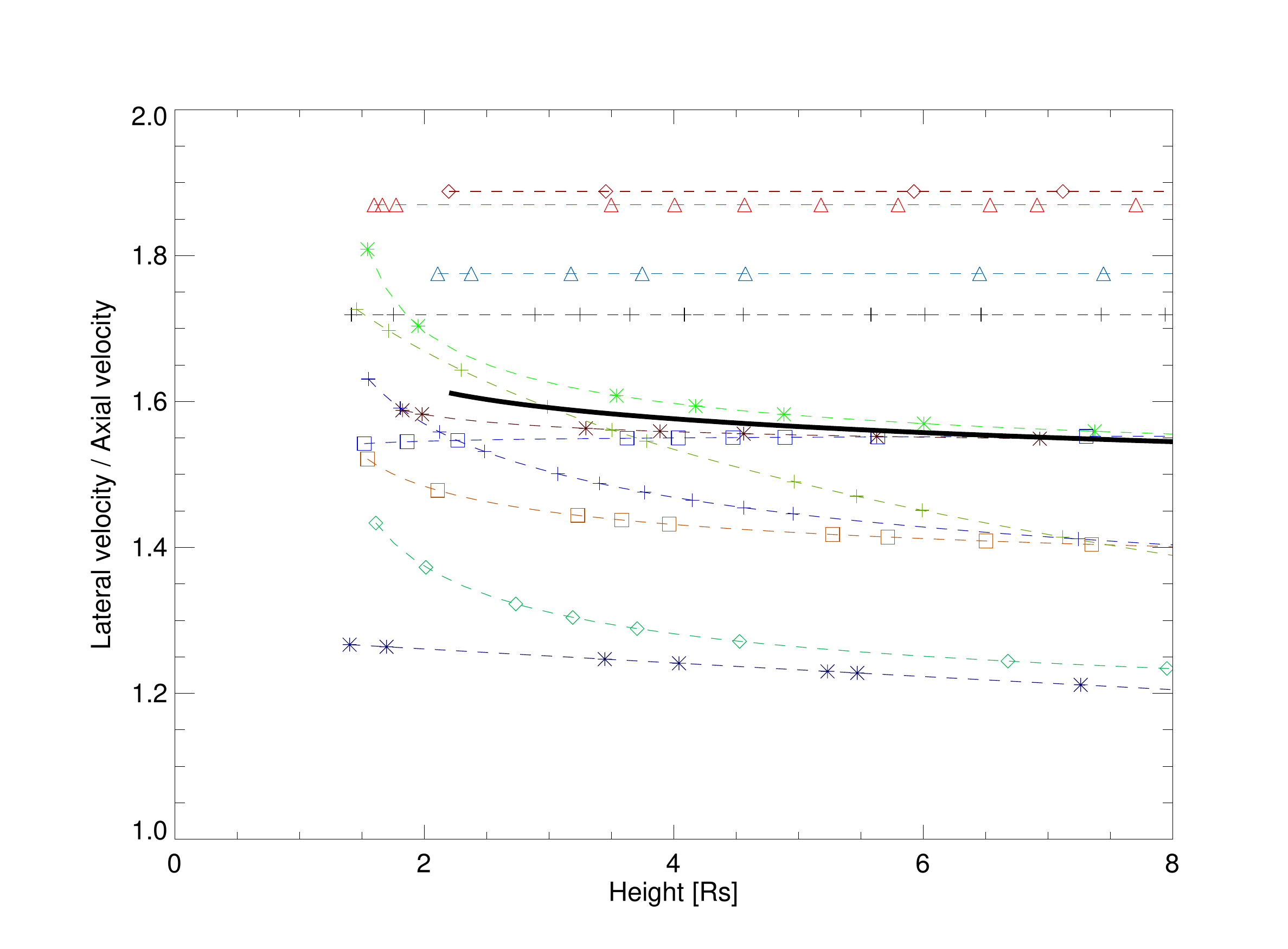}\\
    \vspace{1mm} 
    \includegraphics[width=6.7cm, angle=90, trim=0.7cm 2cm 1.3cm 1.7cm, clip=true]{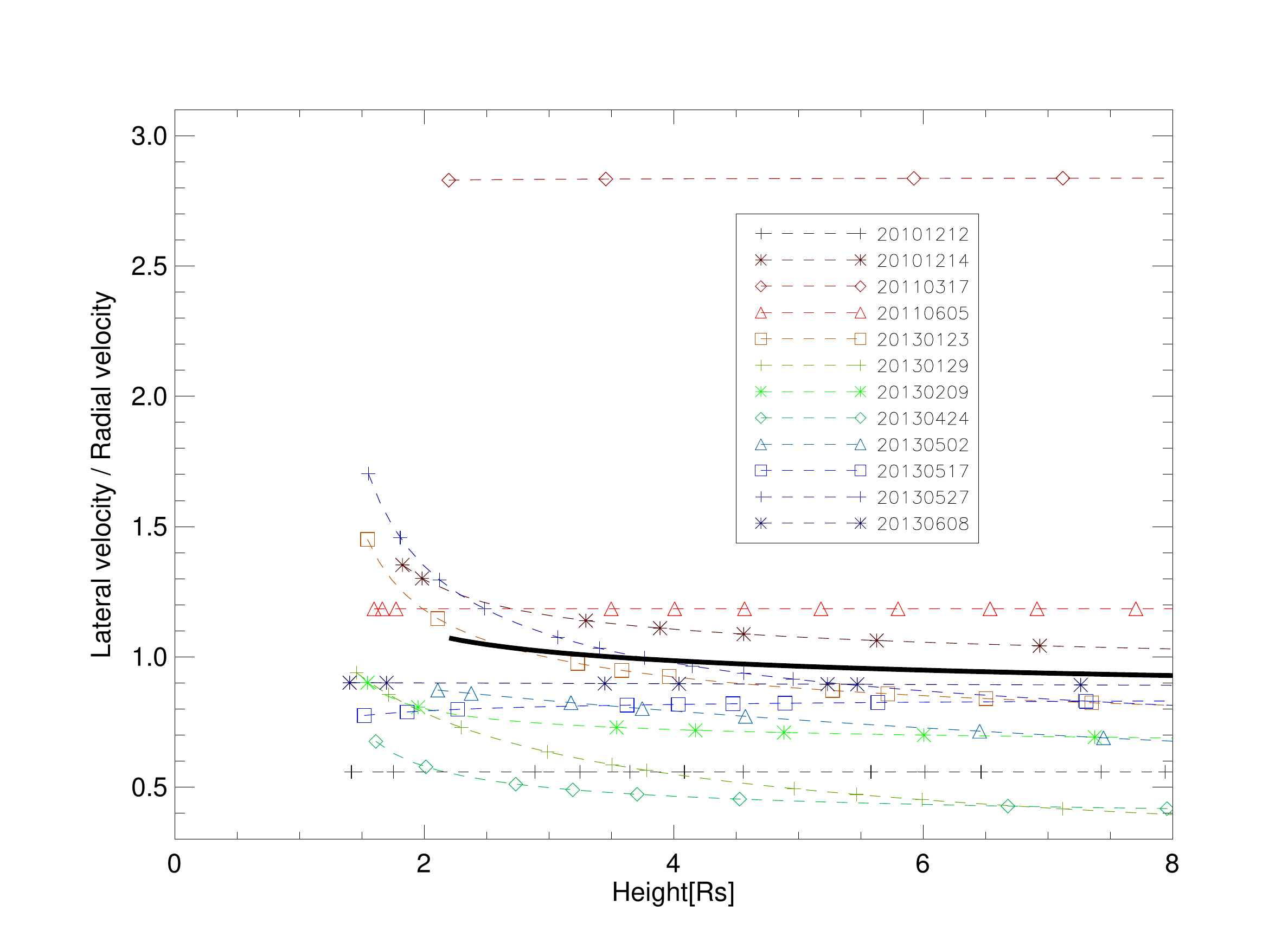}
    \caption{Speed ratios as a function of height. Top panel: Lateral to axial speed. Bottom panel: Lateral to radial (propagation) speed. The dashed colored lines correspond to individual events (see legend in  bottom panel), which also applies to the top plot. The black solid line is their average, starting at the first height where all events could be measured.}
    \label{fig:velocities}
\end{figure}

Expansion speeds in km\,s$^{-1}$ have also been computed along both directions; their values at 6\,\Rsolar~are listed in the last two columns of Table \ref{table}. The top panel of Fig. \ref{fig:velocities} displays the ratio of the speed along the main axes of CMEs (lateral speed) to the perpendicular (axial speed) as a function of height. It should be noted that both speeds are directly proportional at nearly all heights, with the lateral speed being on average $\approx$\,1.6 higher than the axial value (more precisely 1.56\,$\pm$\,0.22). The bottom panel shows the ratio of lateral to radial (propagation) speeds for all analyzed events. In the first few solar radii, some events show a much more height-dependent ratio than others;  after $\sim$\,4\,\Rsolar\ the ratio becomes nearly constant, with the average among all events being close to one (0.96\,$\pm$\,0.63). In general, it can be said that the lateral speeds tend to be higher than the axial and similar to the radial (propagation) speeds, although this is not the case for a few events. 

\begin{figure}
    \centering
    \includegraphics[width=6.7cm, angle=90, trim=0.7cm 2cm 2.0cm 1.7cm, clip=true]{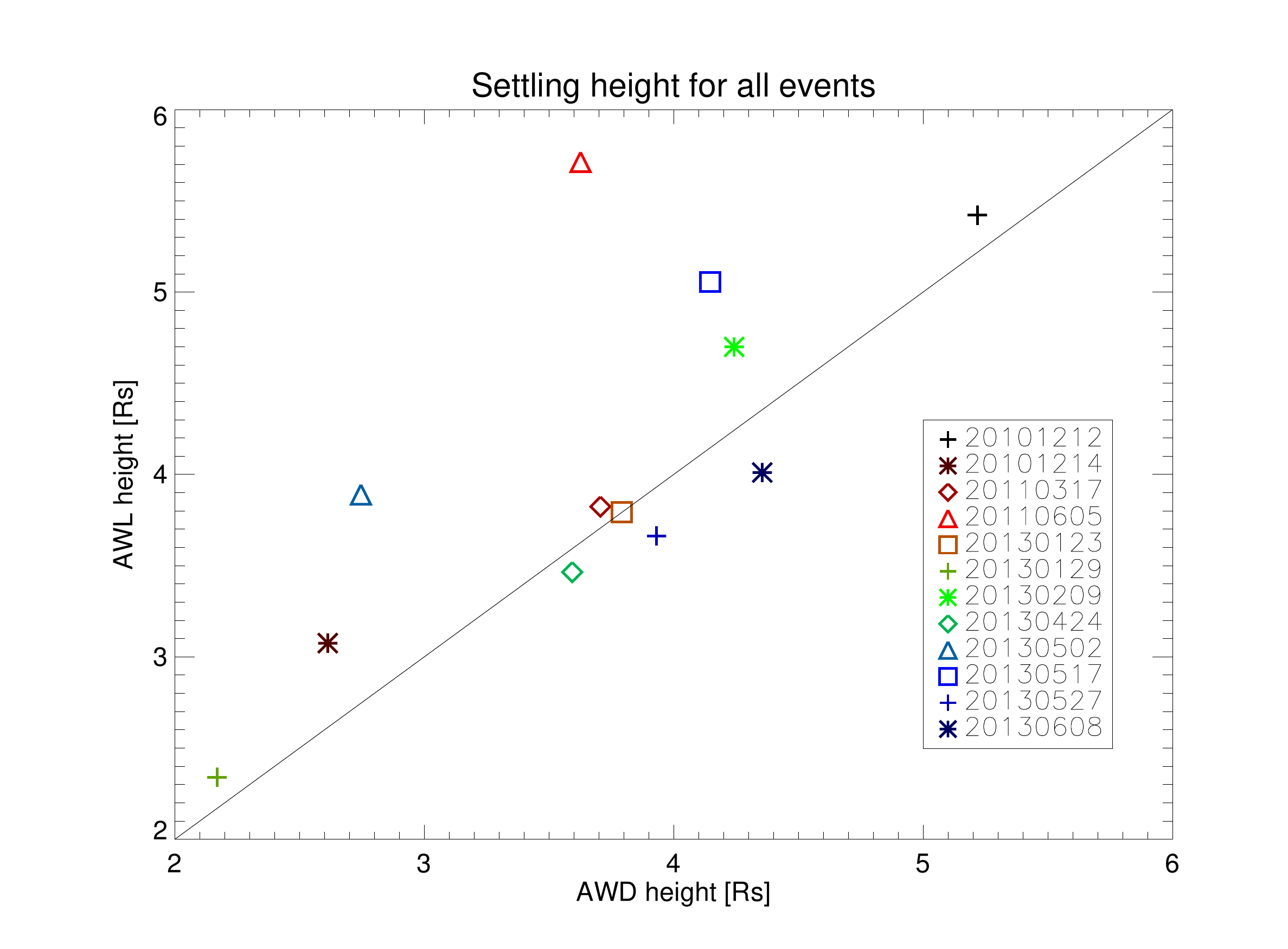}
    \caption{Settling height of $AW_L$ vs. that of $AW_D$ for all analyzed events. The solid line indicates a slope equal to 1.}
    \label{fig:setheight}
\end{figure}

In order to find a means of evaluating the height at which the angular expansion of CMEs stabilizes, we define the settling height as the height at which $AW_L$ and $AW_D$ change less than 5\% per solar radius. The settling heights of $AW_L$ and $AW_D$ are plotted against each other in Fig. \ref{fig:setheight}, where each symbol represents a particular event. In general, events with larger settling heights in one direction tend to have larger settling heights in the other. From the figure, a slight trend toward larger settling heights for $AW_L$ can also be inferred. 

\section{Discussion and conclusions}
\label{s:discussion}

Here we  analyzed twelve polar CME events that took place during epochs of quadrature between spacecraft, and whose eruption could be tracked from their inception in the EUV low corona. The evolution of their morphology was characterized by means of multi-viewpoint coronal observations and a forward-modeling tool that was applied to several time instants. Even though 3D modeling of the morphology of CMEs is a problem with multiple possible solutions, a proper combination of spacecraft location and CME propagation direction allows a better constrained fit. More specifically, conditions for the good assessment of morphological parameters of CMEs are given in this study by (i) the stereoscopic views provided by the STEREO spacecraft located at the ecliptic while away from the Sun-Earth line, plus the view from Earth's perspective provided by SOHO; (ii) a propagation direction of the analyzed CMEs nearly perpendicular to the plane containing these spacecraft; and (iii) the consideration of several time instants for the analysis, including the initial moments in the low corona. 

One of the main findings of this study refers to the amount of asymmetry in the extents and expansion of CMEs in the lateral and axial directions of their embedded flux ropes. In fact, $AW_L$ is greater than $AW_D$ at all heights and for all events. This is imposed by the model, which defines $\alpha$ and $\delta$ as always positive, on the basis of the widths of observed face-on and edge-on CMEs by \citet{Cremades-Bothmer2004}. If any of the analyzed CMEs had a wide flux rope diameter, but a narrow face-on width, i.e., $AW_D>AW_L$, the model would not have been able to successfully fit it.

\begin{figure*}
    \sidecaption
    \centering
    \includegraphics[width=9cm, angle=90, trim=0.7cm 2cm 2.0cm 1.7cm, clip=true]{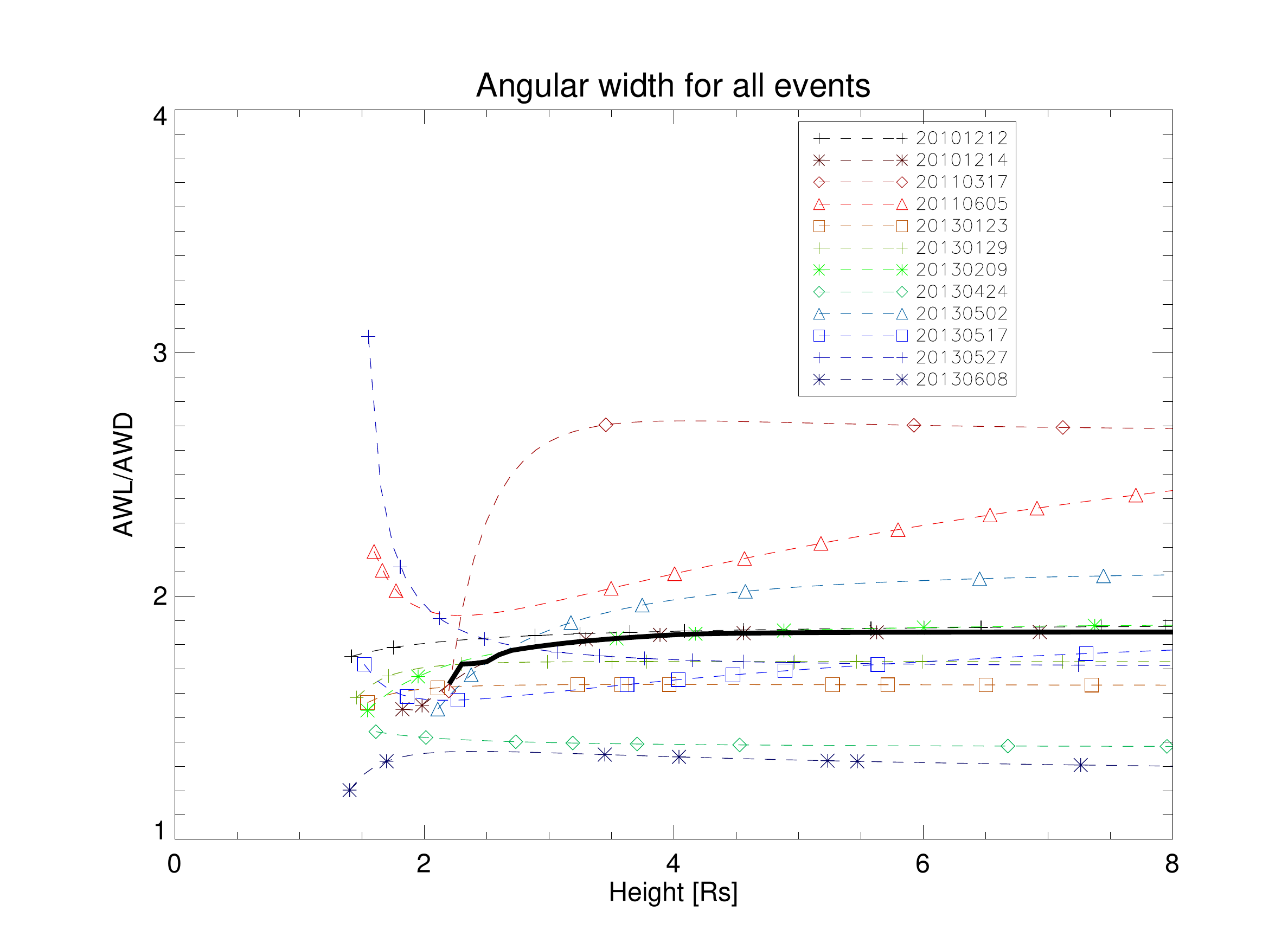}
    \caption{Ratio $AW_L/AW_D$. The dashed colored lines correspond to individual events. The black solid line is their average, starting at the first height where all events could be measured.}
    \label{fig:awloverawd}
\end{figure*}

This can be also visualized in Fig. \ref{fig:awloverawd}, which shows the ratio $AW_L/AW_D$ as a function of height. This ratio presents steeper variations for individual events in the first few solar radii. It then stabilizes at differing values ranging from about 1.3 to 2.7, depending on the event. The average $AW_L/AW_D$ ratio over height for all events is 1.84\,$\pm$\,0.37, and is represented in the figure by the black solid line. Different angular widths depending on the flux rope orientation with respect to the observer had been suggested by \citet{StCyr-etal2004} and \citet{Cremades-Bothmer2005}, who found a ratio $AW_L/AW_D\approx$\,1.6. We would like to note that the latter study measured widths of specific features --circular void and prominence extent-- rather than full AWs. Moreover, these ratios arose from the comparison of the average lateral (broadside, $AW_L$) and axial ($AW_D$) widths found for different events from the perspective of SOHO, the single available view at the time. The first case study for which it was possible to determine the $AW_L/AW_D$ ratio for the same CME was presented by \citet{Cabello-etal2016}, who found a value $\approx$\,1.6 as well. The analysis of \citet{Krall-StCyr2006} performed on a parameterized model CME yielded a ratio of 1.9 for the best considered fit.

Although significant work has been devoted to understanding how flux-rope cross sections evolve, little has been reported on the behavior of the lateral versus axial extents. In the case study by \citet{Wood-etal2010} the CME flux rope was interpreted to be very fat at early times, and to become narrower and flatter as it evolved outward from the Sun. \citet{Janvier-etal2014} discuss the unfeasibility of comparing distributions of CME apparent AWs and of the radius of magnetic clouds, given that CME apparent widths involve projection and selection effects that hinder the proper measurement of the extension of a flux rope's axis and radius. \citet{Balmaceda-etal2018}, in turn, classified CMEs according to their morphological appearance, where the L (loop) and F (flux rope) types are attributed to CMEs seen face-on and edge-on, respectively, and found an average AW of L-CMEs nearly two times larger than that of F-CMEs.

Unlike the lateral versus axial asymmetry, the aspect ratio of CMEs, defined as the quotient of the CME center height over the CME's half width (center of the flux rope/diameter), has been addressed by a fair number of studies in the literature. For instance, \citet{Krall-Chen2001} noticed that when looplike features could be measured in EIT, the loop expanded in width faster than it moved outward, implying a decreasing aspect ratio at low heights. \citet{Patsourakos-etal2010} analyzed the  3D expansion of a CME and found two distinct phases, with the first of them also exhibiting smaller aspect ratios and the following phase showing an almost constant aspect ratio. A constant aspect ratio is commonly regarded as a sign of self-similarity, meaning that CMEs maintain their shape as they expand \citep{Wood-etal2009}. \citet{Veronig-etal2018} also find a fast decrease in the CME aspect ratio in the first stages of the CME genesis. A similar concept of aspect ratio has been investigated much farther in the heliosphere, in relation with the cross section of magnetic clouds regarded as cylindrical objects. It has been proposed that these cross sections are greatly flattened \citep[e.g.,][]{Riley-Crooker2004, Nieves-Chin-etal2018}. However the ``pancaking'' may cease at some point before the arrival of these interplanetary structures to Earth \citep{Savani-etal2011}.

We  also find that the early times of the analyzed eruptions are the most important in terms of morphology, as demonstrated by the steep slope of the aspect ratio and of the difference between the AW change rates in the first few solar radii. This result is consistent with previous findings, like those of \citet{Zhang-etal2004} and \citet{Patsourakos-etal2010}. The former noticed a superexpansion in the lateral direction during the acceleration phase in the lower corona, in contrast to the self-similar expansion pattern of constant AW observed in the outer corona. \citet{Patsourakos-etal2010} also proposed two phases of evolution, with the initial one exhibiting a faster expansion and rise below 2\,\Rsolar, followed by a nearly constant aspect ratio implying a self-similar evolution. Other observational studies revealing that the first instants of CME evolution present rapid changes and are decisive in their morphological development, report the acceleration of most events peaking at heights below 0.5\,\Rsolar\ \citep{Bein-etal2011} or 2\,\Rsolar\ \citep{Joshi-Srivastava2011}. 

The main results arising from the multi-viewpoint analysis of these 12 high-latitude CMEs can be summarized as follows:
\begin{itemize}
    \item Angular widths along the main axes of CMEs ($AW_L$, face-on view) are larger than angular widths in the orthogonal direction ($AW_D$, edge-on view) by an average factor of $\approx1.8$ (Figs. \ref{fig:bothAWs} and \ref{fig:awloverawd}).
    \item Both angular widths change considerably with height below $\sim$\,3\,\Rsolar\ (Fig. \ref{fig:bothAWs}).
    \item The growth rate of $AW_L$ is higher than that of $AW_D$ below $\sim$\,3\,\Rsolar\ (Fig. \ref{fig:difference}).
    \item The ratio of the two  expansion speeds, namely in the lateral ($AW_L$) and axial ($AW_D$) directions, is nearly constant after $\sim$\,4\,\Rsolar\, with an average $\approx$\,1.6 (Fig. \ref{fig:velocities}, top). 
    \item On average, the expansion speed in the lateral ($AW_L$) direction is similar to the outward radial propagation speed (Fig. \ref{fig:velocities}, bottom).
    \item The settling heights of $AW_L$ tend to be larger than those of $AW_D$, showing some degree of proportionality.
\end{itemize}

According to the obtained results, we conclude that the expansion of the studied CMEs below $\sim$\,3\,\Rsolar\ cannot be considered self-similar and is evidently  asymmetric, with higher expansion speeds in the lateral direction. This result has important implications for understanding and modeling not only CME morphology, but also shock generation, EUV wave formation, and their properties, such as profile and shape, among others.

As a follow-up to this work, we plan to understand our findings by means of MHD-based models. Although most of them do not consider the asymmetric expansion of CMEs and of their embedded flux ropes, they do address the mechanisms governing the overall expansion. For instance, \citet{Byrne-etal2010} attribute the expansion of a modeled CME from 30$^{\circ}$ to 60$^{\circ}$ over a height range of 2\,--\,46\,\Rsolar\ to the convection with the ambient solar wind in a diverging geometry, and to the pressure gradient between the flux rope and solar wind. \citet{Mishra-Wang2018} treat CMEs as axisymmetric cylinders assuming self-similar expansion, with the net force direction in agreement with the expansion acceleration, suggesting that the thermal pressure force is the internal driver of CME expansion and the Lorentz force the restraining agent. \citet{Liu-etal2008} also found that that  thermal pressure accounts for most of the acceleration of a CME, with the magnetic pressure contributing only to the acceleration early in the evolution up to $\sim$\,3\,\Rsolar. In turn, \citet{Patsourakos-etal2010} argue that the initial overexpansion of an erupting flux rope is caused neither by the decreasing ambient pressure as the flux rope rises nor by photospheric motions. Instead, they suggest that it results from an expansion of the flux surfaces of the poloidal flux external to the flux rope itself (driven by flux conservation) and/or from the rapid addition of flux by reconnection in a current sheet under a growing pre-existing flux rope in growth or a new one in formation. Expansion in the lateral direction is not addressed in this interpretation.

Out-of-the-ecliptic missions like Solar Orbiter will enable similar analyses that minimize uncertainties due to projection effects, but on events propagating close to the ecliptic plane, particularly Earth-directed. All currently existing coronagraphs lie on this plane, thus hindering proper 3D reconstruction of potentially geoeffective CMEs.

\begin{acknowledgements}
HC is member of the Carrera del Investigador Cient\'ifico (CONICET). FAI is a postdoctoral fellow of CONICET and LAM holds a scholarship from the National Inter-University Council (Argentina). The authors are grateful to an anonymous referee for useful comments and suggestions. Authors appreciate financial support from UTN grants UTI4915TC and ME5445 and  acknowledge use of data from the STEREO (NASA), SDO (NASA), and SOHO (ESA/NASA) missions, produced by the SECCHI, AIA, and LASCO international consortia. This work uses data from the SOHO/LASCO CME Catalog generated and maintained at the CDAW Data Center by NASA and the CUA in cooperation with NRL.
\end{acknowledgements}

%
   \bibliographystyle{aa} 
   \bibliography{fullbiblio.bib} 

\begin{thebibliography}{43}
\expandafter\ifx\csname natexlab\endcsname\relax\def\natexlab#1{#1}\fi

\bibitem[{{Balmaceda} {et~al.}(2018){Balmaceda}, {Vourlidas}, {Stenborg}, \&
  {Dal Lago}}]{Balmaceda-etal2018}
{Balmaceda}, L.~A., {Vourlidas}, A., {Stenborg}, G., \& {Dal Lago}, A. 2018,
  \apj, 863, 57

\bibitem[{{Bein} {et~al.}(2011){Bein}, {Berkebile-Stoiser}, {Veronig},
  {Temmer}, {Muhr}, {Kienreich}, {Utz}, \& {Vr{\v{s}}nak}}]{Bein-etal2011}
{Bein}, B.~M., {Berkebile-Stoiser}, S., {Veronig}, A.~M., {et~al.} 2011, \apj,
  738, 191

\bibitem[{{Bothmer} \& {Schwenn}(1998)}]{Bothmer-Schwenn1998}
{Bothmer}, V. \& {Schwenn}, R. 1998, Annales Geophysicae, 16, 1

\bibitem[{{Brueckner} {et~al.}(1995){Brueckner}, {Howard}, {Koomen},
  {Korendyke}, {Michels}, {Moses}, {Socker}, {Dere}, {Lamy}, {Llebaria},
  {Bout}, {Schwenn}, {Simnett}, {Bedford}, \& {Eyles}}]{Brueckner-etal1995}
{Brueckner}, G.~E., {Howard}, R.~A., {Koomen}, M.~J., {et~al.} 1995, Solar
  Physics, 162, 357

\bibitem[{{Byrne} {et~al.}(2010){Byrne}, {Maloney}, {McAteer}, {Refojo}, \&
  {Gallagher}}]{Byrne-etal2010}
{Byrne}, J.~P., {Maloney}, S.~A., {McAteer}, R.~T.~J., {Refojo}, J.~M., \&
  {Gallagher}, P.~T. 2010, Nature Communications, 1, 74

\bibitem[{{Cabello} {et~al.}(2016){Cabello}, {Cremades}, {Balmaceda}, \&
  {Dohmen}}]{Cabello-etal2016}
{Cabello}, I., {Cremades}, H., {Balmaceda}, L., \& {Dohmen}, I. 2016, Solar
  Physics, 291, 1799

\bibitem[{{Cremades} \& {Bothmer}(2004)}]{Cremades-Bothmer2004}
{Cremades}, H. \& {Bothmer}, V. 2004, A\&A, 422, 307

\bibitem[{{Cremades} \& {Bothmer}(2005)}]{Cremades-Bothmer2005}
{Cremades}, H. \& {Bothmer}, V. 2005, in IAU Symp., Vol. 226, Coronal and
  Stellar Mass Ejections, ed. K.~{Dere}, J.~{Wang}, \& Y.~{Yan}, 48--54

\bibitem[{{Crifo} {et~al.}(1983){Crifo}, {Picat}, \&
  {Cailloux}}]{Crifo-etal1983}
{Crifo}, F., {Picat}, J.~P., \& {Cailloux}, M. 1983, Solar Physics, 83, 143

\bibitem[{{Dere} {et~al.}(1999){Dere}, {Brueckner}, {Howard}, {Michels}, \&
  {Delaboudiniere}}]{Dere-etal1999}
{Dere}, K.~P., {Brueckner}, G.~E., {Howard}, R.~A., {Michels}, D.~J., \&
  {Delaboudiniere}, J.~P. 1999, ApJ, 516, 465

\bibitem[{{Domingo} {et~al.}(1995){Domingo}, {Fleck}, \&
  {Poland}}]{Domingo-etal1995}
{Domingo}, V., {Fleck}, B., \& {Poland}, A.~I. 1995, Solar Physics, 162, 1

\bibitem[{{Gibson} {et~al.}(2006){Gibson}, {Foster}, {Burkepile}, {de Toma}, \&
  {Stanger}}]{Gibson-etal2006}
{Gibson}, S.~E., {Foster}, D., {Burkepile}, J., {de Toma}, G., \& {Stanger}, A.
  2006, \apj, 641, 590

\bibitem[{{Howard} {et~al.}(1982){Howard}, {Michels}, {Sheeley}, \&
  {Koomen}}]{Howard-etal1982}
{Howard}, R.~A., {Michels}, D.~J., {Sheeley}, N.~R., \& {Koomen}, M.~J. 1982,
  \apjl, 263, L101

\bibitem[{{Howard} {et~al.}(2008){Howard}, {Moses}, {Vourlidas}, {Newmark},
  {Socker}, {Plunkett}, {Korendyke}, {Cook}, {Hurley}, {Davila}, \& {et
  al.}}]{Howard-etal2008}
{Howard}, R.~A., {Moses}, J.~D., {Vourlidas}, A., {et~al.} 2008, Space Sci.
  Rev., 136, 67

\bibitem[{{Janvier} {et~al.}(2014){Janvier}, {D{\'e}moulin}, \&
  {Dasso}}]{Janvier-etal2014}
{Janvier}, M., {D{\'e}moulin}, P., \& {Dasso}, S. 2014, \solphys, 289, 2633

\bibitem[{{Joshi} \& {Srivastava}(2011)}]{Joshi-Srivastava2011}
{Joshi}, A.~D. \& {Srivastava}, N. 2011, \apj, 739, 8

\bibitem[{{Kaiser} {et~al.}(2008){Kaiser}, {Kucera}, {Davila}, {St.~Cyr},
  {Guhathakurta}, \& {Christian}}]{Kaiser-etal2008}
{Kaiser}, M.~L., {Kucera}, T.~A., {Davila}, J.~M., {et~al.} 2008, Space Sci.
  Rev., 136, 5

\bibitem[{{Krall} {et~al.}(2001){Krall}, {Chen}, {Duffin}, {Howard}, \&
  {Thompson}}]{Krall-Chen2001}
{Krall}, J., {Chen}, J., {Duffin}, R.~T., {Howard}, R.~A., \& {Thompson}, B.~J.
  2001, \apj, 562, 1045

\bibitem[{{Krall} \& {St. Cyr}(2006)}]{Krall-StCyr2006}
{Krall}, J. \& {St. Cyr}, O.~C. 2006, \apj, 652, 1740

\bibitem[{{Lemen} {et~al.}(2012){Lemen}, {Title}, {Akin}, {Boerner}, {Chou},
  {Drake}, {Duncan}, {Edwards}, {Friedlaender}, {Heyman}, \& {et
  al.}}]{Lemen-etal2012}
{Lemen}, J.~R., {Title}, A.~M., {Akin}, D.~J., {et~al.} 2012, Solar Physics,
  275, 17

\bibitem[{{Liu} {et~al.}(2008){Liu}, {Opher}, {Cohen}, {Liewer}, \&
  {Gombosi}}]{Liu-etal2008}
{Liu}, Y.~C.~M., {Opher}, M., {Cohen}, O., {Liewer}, P.~C., \& {Gombosi}, T.~I.
  2008, The Astrophysical Journal, 680, 757

\bibitem[{{Marubashi} {et~al.}(2015){Marubashi}, {Akiyama}, {Yashiro},
  {Gopalswamy}, {Cho}, \& {Park}}]{Marubashi-etal2015}
{Marubashi}, K., {Akiyama}, S., {Yashiro}, S., {et~al.} 2015, Solar Physics,
  290, 1371

\bibitem[{{Mierla} {et~al.}(2009){Mierla}, {Inhester}, {Marqu{\'e}},
  {Rodriguez}, {Gissot}, {Zhukov}, {Berghmans}, \& {Davila}}]{Mierla-etal2009}
{Mierla}, M., {Inhester}, B., {Marqu{\'e}}, C., {et~al.} 2009, Solar Physics,
  259, 123

\bibitem[{{Mishra} \& {Wang}(2018)}]{Mishra-Wang2018}
{Mishra}, W. \& {Wang}, Y. 2018, The Astrophysical Journal, 865, 50

\bibitem[{{Moran} \& {Davila}(2004)}]{Moran-Davila2004}
{Moran}, T.~G. \& {Davila}, J.~M. 2004, Science, 305, 66

\bibitem[{{Nieves-Chinchilla} {et~al.}(2018){Nieves-Chinchilla}, {Linton},
  {Hidalgo}, \& {Vourlidas}}]{Nieves-Chin-etal2018}
{Nieves-Chinchilla}, T., {Linton}, M.~G., {Hidalgo}, M.~A., \& {Vourlidas}, A.
  2018, \apj, 861, 139

\bibitem[{{Patsourakos} {et~al.}(2010){Patsourakos}, {Vourlidas}, \&
  {Kliem}}]{Patsourakos-etal2010}
{Patsourakos}, S., {Vourlidas}, A., \& {Kliem}, B. 2010, \aap, 522, A100

\bibitem[{{Pesnell} {et~al.}(2012){Pesnell}, {Thompson}, \&
  {Chamberlin}}]{Pesnell-etal2012}
{Pesnell}, W.~D., {Thompson}, B.~J., \& {Chamberlin}, P.~C. 2012, Solar
  Physics, 275, 3

\bibitem[{{Riley} \& {Crooker}(2004)}]{Riley-Crooker2004}
{Riley}, P. \& {Crooker}, N.~U. 2004, \apj, 600, 1035

\bibitem[{{Savani} {et~al.}(2011){Savani}, {Owens}, {Rouillard}, {Forsyth},
  {Kusano}, {Shiota}, \& {Kataoka}}]{Savani-etal2011}
{Savani}, N.~P., {Owens}, M.~J., {Rouillard}, A.~P., {et~al.} 2011, \apj, 731,
  109

\bibitem[{{Savani} {et~al.}(2015){Savani}, {Vourlidas}, {Szabo}, {Mays},
  {Richardson}, {Thompson}, {Pulkkinen}, {Evans}, \&
  {Nieves-Chinchilla}}]{Savani-etal2015}
{Savani}, N.~P., {Vourlidas}, A., {Szabo}, A., {et~al.} 2015, Space Weather,
  13, 374

\bibitem[{{St. Cyr} {et~al.}(2004){St. Cyr}, {Cremades}, {Bothmer}, {Krall}, \&
  {Burkepile}}]{StCyr-etal2004}
{St. Cyr}, O.~C., {Cremades}, H., {Bothmer}, V., {Krall}, J., \& {Burkepile},
  J.~T. 2004, in AGU Fall Meeting Abstracts, Vol. 2004, SH22A--04

\bibitem[{{Thernisien} {et~al.}(2009){Thernisien}, {Vourlidas}, \&
  {Howard}}]{Thernisien-etal2009}
{Thernisien}, A., {Vourlidas}, A., \& {Howard}, R.~A. 2009, Solar Physics, 256,
  111

\bibitem[{{Thernisien} {et~al.}(2011){Thernisien}, {Vourlidas}, \&
  {Howard}}]{Thernisien-etal2011}
{Thernisien}, A., {Vourlidas}, A., \& {Howard}, R.~A. 2011, J. Atmos.
  Solar-Terr. Phys., 73, 1156

\bibitem[{{Thernisien} {et~al.}(2006){Thernisien}, {Howard}, \&
  {Vourlidas}}]{Thernisien-etal2006}
{Thernisien}, A.~F.~R., {Howard}, R.~A., \& {Vourlidas}, A. 2006, \apj, 652,
  763

\bibitem[{{Veronig} {et~al.}(2018){Veronig}, {Podladchikova}, {Dissauer},
  {Temmer}, {Seaton}, {Long}, {Guo}, {Vr{\v{s}}nak}, {Harra}, \&
  {Kliem}}]{Veronig-etal2018}
{Veronig}, A.~M., {Podladchikova}, T., {Dissauer}, K., {et~al.} 2018, \apj,
  868, 107

\bibitem[{{Vourlidas} {et~al.}(2013){Vourlidas}, {Lynch}, {Howard}, \&
  {Li}}]{Vourlidas-etal2013}
{Vourlidas}, A., {Lynch}, B.~J., {Howard}, R.~A., \& {Li}, Y. 2013, Solar
  Physics, 284, 179

\bibitem[{{Wood} {et~al.}(2010){Wood}, {Howard}, \& {Socker}}]{Wood-etal2010}
{Wood}, B.~E., {Howard}, R.~A., \& {Socker}, D.~G. 2010, \apj, 715, 1524

\bibitem[{{Wood} {et~al.}(2009){Wood}, {Howard}, {Thernisien}, {Plunkett}, \&
  {Socker}}]{Wood-etal2009}
{Wood}, B.~E., {Howard}, R.~A., {Thernisien}, A., {Plunkett}, S.~P., \&
  {Socker}, D.~G. 2009, \solphys, 259, 163

\bibitem[{{Wood} {et~al.}(1999){Wood}, {Karovska}, {Chen}, {Brueckner}, {Cook},
  \& {Howard}}]{Wood-etal1999}
{Wood}, B.~E., {Karovska}, M., {Chen}, J., {et~al.} 1999, \apj, 512, 484

\bibitem[{{Yashiro} {et~al.}(2004){Yashiro}, {Gopalswamy}, {Michalek},
  {St.~Cyr}, {Plunkett}, {Rich}, \& {Howard}}]{Yashiro-etal2004}
{Yashiro}, S., {Gopalswamy}, N., {Michalek}, G., {et~al.} 2004, \jgr~(Space
  Physics), 109, 7105

\bibitem[{{Yurchyshyn} {et~al.}(2007){Yurchyshyn}, {Hu}, {Lepping}, {Lynch}, \&
  {Krall}}]{Yurchyshyn-etal2007}
{Yurchyshyn}, V., {Hu}, Q., {Lepping}, R.~P., {Lynch}, B.~J., \& {Krall}, J.
  2007, Advances in Space Research, 40, 1821

\bibitem[{{Zhang} {et~al.}(2004){Zhang}, {Dere}, {Howard}, \&
  {Vourlidas}}]{Zhang-etal2004}
{Zhang}, J., {Dere}, K.~P., {Howard}, R.~A., \& {Vourlidas}, A. 2004, \apj,
  604, 420

\end{thebibliography}
%

%
\end{document}